%% file: main.tex
\documentclass[conference,a4paper,10pt,times]{IEEEtran}

\usepackage{hyperref}
\usepackage{color, colortbl}

 \usepackage[table]{xcolor}

\usepackage{algorithm}

\usepackage{algorithmic}
\usepackage{array}
\usepackage{amssymb}
\usepackage{subfloat}
\usepackage{inputenc}
\usepackage{censor}
\usepackage{bbding}
\usepackage{array}
\usepackage{amssymb}
\usepackage{arydshln}
\usepackage{amsthm}
\usepackage{amsmath}
\usepackage{fontenc}
\usepackage{textcomp}
\usepackage{scrhack}
\usepackage{xspace}
\usepackage{subfig}
\usepackage{tabulary}

\usepackage{longtable}
\usepackage{tcolorbox}
\usepackage{colortbl}
\usepackage{listings}

\newcommand{\redactedOS}{\texttt{RedactedOS}\xspace}
\newcommand{\uArmor}{$\mu$Armor\xspace}
\newcommand{\uESP}{\texttt{$\mu$ESP}\xspace}
\newcommand{\uScramble}{\texttt{$\mu$Scramble}\xspace}
\newcommand{\uSSP}{\texttt{$\mu$SSP}\xspace}
\newcommand{\uRNG}{\texttt{$\mu$RNG}\xspace}

\usepackage{xcolor}
\usepackage{listings} 
\newcolumntype{K}[1]{>{\centering\arraybackslash}p{#1}}
\usepackage{arydshln}

\usepackage{tabularx}
\definecolor{LRed}{rgb}{1,.8,.8}
\definecolor{MRed}{rgb}{1,.6,.6}
\definecolor{HRed}{rgb}{1,.2,.2}
\definecolor{light-gray}{gray}{0.95}
\definecolor{orange}{cmyk}{0,0.5,1,0}
\definecolor{maroon}{cmyk}{0,0.87,0.68,0.32}
\definecolor{almond}{rgb}{0.94, 0.87, 0.8}
\usepackage[table]{xcolor} 
 \rowcolors{2}{gray!25}{white}

\definecolor{bleudefrance}{rgb}{0.19, 0.55, 0.91}              

\usepackage{url}

\begin{document}

\title{Challenges in Designing Exploit Mitigations for Deeply Embedded Systems}

\author{\IEEEauthorblockN{Ali Abbasi\IEEEauthorrefmark{1},
		Jos Wetzels\IEEEauthorrefmark{2}, Thorsten Holz\IEEEauthorrefmark{1} and
		Sandro Etalle\IEEEauthorrefmark{2}}

	\IEEEauthorblockA{\IEEEauthorrefmark{1} Ruhr-University Bochum,
		Germany\\
		Email: ali.abbasi-i4q@rub.de, thorsten.holz@rub.de}

	\IEEEauthorblockA{\IEEEauthorrefmark{2} Eindhoven University of Technology, The Netherlands\\
	Email: a.l.g.m.wetzels@student.tue.nl, s.etalle@tue.nl
}

}

\maketitle

\maketitle

\begin{abstract}
Memory corruption vulnerabilities have been around for decades and rank among the most prevalent vulnerabilities in embedded systems. Yet this constrained environment poses unique design and implementation challenges that significantly complicate the adoption of common hardening techniques. Combined with the irregular and involved nature of embedded patch management, this results in prolonged vulnerability exposure windows and vulnerabilities that are relatively easy to exploit. Considering the sensitive and critical nature of many embedded systems, this situation merits significant improvement.

In this work, we present the first quantitative study of exploit mitigation adoption in 42 embedded operating systems, showing the embedded world to significantly lag behind the general-purpose world. To improve the security of deeply embedded systems, we subsequently present $\mu$Armor, an approach to address some of the key gaps identified in our quantitative analysis. $\mu$Armor raises the bar for exploitation of embedded memory corruption vulnerabilities, while being adoptable on the short term without incurring prohibitive extra performance or storage costs.
\end{abstract}

\IEEEpeerreviewmaketitle

\input{intro}

\input{background}
\input{quantitative}

\input{design}

\input{implementation}

\input{discussion}

\input{conclusion}

\bibliographystyle{IEEEtran}
\def\UrlBreaks{\do\-}
\bibliography{ref}

\end{document}

%% file: intro.tex
\section{Introduction}\label{sec:intro}

Embedded systems are everywhere, from simple networking equipment to satellite systems. With the rise of cyber-physical systems and the \emph{Internet of Things} (IoT), these systems are set to proliferate throughout all aspects of everyday life. Due to their ubiquitous and often sensitive and critical nature, embedded systems pose many security and privacy concerns. Unfortunately, proper attention to security in the embedded world tends to be scarce in practice. This tendency becomes clear from various studies~\cite{automotive_analysis,costin_firmwares} revealing security flaws in a wide variety of embedded systems.
These flaws are far from hypothetical. One example is the high-profile attack on embedded systems known for having been used to construct the IoT-powered botnet Mirai~\cite{mirai}. Yet embedded systems security is seen as lagging behind what we have come to expect of our general purpose (e.g., desktop and server) systems~\cite{koopman_embedded_system_security:2004}. 

Embedded \emph{binary security}, in particular, is an area where exploitation of vulnerabilities is significantly easier than on general-purpose systems. This is exemplified by a recent incident where a previously unknown group calling themselves the \emph{Shadow Brokers} released a cache of exploits which they claimed belonged to the supposedly state-sponsored \emph{Equation Group}~\cite{kaspersky_equation:2015} threat actor. Among this cache was a set of exploits for high-end firewall equipment, none of which had to bypass any exploit mitigations. Despite these threats, and the general perception of embedded systems binary security as lagging, we do not yet have a clear understanding of the existing gap in security technologies available for embedded systems compared to general-purpose computers. 

In this paper, we study this problem and focus on three major bare-minimum and well-known exploit mitigation techniques. More specifically, we focus on \emph{Executable Space Protection} (ESP, also known as NX, DEP, or $W \oplus X$ policy), \emph{Address Space Layout Randomization} (ASLR) and \emph{stack canaries} to identify the gap in security technologies specifically in the memory corruption and exploitation domain. These exploit mitigation methods are almost universally available on general-purpose computers. We investigate whether such mitigations are available in 42 major embedded Operating Systems (OS). We found that half of them provide such mitigations and the majority of them belong to the group of so-called high-end embedded OSes. However, when it comes to lower-end embedded OSes that are mostly being used in so-called \emph{deeply embedded systems}, exploit mitigations are almost absent. The \emph{deeply embedded systems} as described by Koopman et al.~\cite{koopman_deep_survivability:2007} are a subset of embedded systems which usually rely on 8-, 16- or (at the higher end of the spectrum) 32-bit micro-controllers. These systems tend to come in the form of a micro-controller unit (MCU) or more extensive System-on-Chip (SoC) devices, embedding both the core as well as memory and peripheral devices into a single chip.

We investigated 78 common~\cite{ubm_study} embedded microprocessors and microcontrollers to understand whether this universal lack of exploit mitigations is caused by lack of hardware feature support (i.e., Memory Management Unit (MMU), Memory Protection Unit (MPU), or Hardware-supported ESP).  

While the lack of exploit mitigations for bare-metal embedded systems (i.e., devices without an OS) is addressed in recent research~\cite{clements2017protecting}, no research suggests a solution for deeply embedded systems that are running multi-stack, multi-threaded and real-time capable operating systems. This lack of solutions comes with risks that are expected to increase. According to VDC Research~\cite{vdcresearch}, the IoT has caused developers for deeply embedded systems to move away from bare-metal embedded systems towards deeply embedded systems running an OS.

Based on our quantitative analysis on the lack of exploit mitigation support on deeply embedded OSes and considering the recent research by Celements et al.~\cite{clements2017protecting} which addresses bare-metal deeply-embedded systems, we introduce \uArmor, a set of LLVM passes that harden deeply embedded systems that are running an OS. Our goal is to bring exploit mitigation baselines from general-purpose computers to the most constrained end of the embedded spectrum. More specifically, \uArmor adds several exploit mitigation strategies to such systems. We have built a prototype implementation of this concept and evaluate the overhead imposed by \uArmor in terms of code size, data size, memory usage, and runtime overhead. Our empirical results indicate that the induced overhead is very low and suitable even for deeply embedded systems.

\smallskip \noindent
In summary, we make the following contributions:
\begin{itemize}
	
\item We perform a comprehensive, quantitative study of exploit mitigation adoption in 42 embedded operating systems. This analysis clearly shows that embedded systems severely lag behind general purpose ones.

\item We perform a systematic identification of the challenges faced by embedded exploit mitigation adoption efforts. Based on this analysis, we identify two major open problems and subsequently introduce a solution to address them.

\item We propose, implement and evaluate \uArmor, an exploit mitigation baseline design for deeply embedded systems running a multi-stack, multi-threaded, real-time capable operating system. Additionally, as part of \uArmor, we present \uSSP, a stack canary scheme with a modular violation policy handler allowing for the preservation of the system's availability. 
	
\end{itemize}

%% file: background.tex
\section{Mitigations Baseline}\label{backgroundsection}

The term \emph{embedded systems} covers a large number of different devices that are dedicated to a specific purpose. In this work, we focus on so-called \emph{deeply embedded systems}~\cite{koopman_deep_survivability:2007}, a subset of embedded systems which usually rely on 8-, 16- or (at the higher end of the spectrum) 32-bit micro-controllers. These systems tend to come in the form of a micro-controller unit (MCU) or more extensive System-on-Chip (SoC) devices, embedding both the core as well as memory and peripheral devices into a single chip. Such a high level of integration allows for low production cost and simplifies production, but at the same time constrains capabilities with regards to memory size, speed and power consumption. 
Deeply embedded systems often lack user interfaces or have uncommon ones and tend to run extremely minimal OSes (often with real-time capabilities) or no OS at all (bare-metal). 

We are interested in defining the absolute minimum set of mitigations which should be reasonably expected to be present in all modern embedded systems. Because of the sheer diversity of embedded systems, we do not select our baseline by strict criteria. Instead, we require them to be adaptable across the embedded spectrum and not rely on any specialized hardware feature not commonly present in Commercial off-the-shelf (COTS) embedded hardware. 
The minimum embedded exploit mitigation baseline we selected comprised of ESP,
ASLR, and stack canaries. These mitigations were selected because they are complementary and have been integrated into virtually all modern general-purpose OSes and development toolchains, including those widely used in the embedded world.  As such they are well understood and can reasonably be considered to be the absolute minimum in modern exploit mitigations.

%% file: quantitative.tex
\section{Quantitative Analysis of Exploit Mitigation Methods in Embedded Systems}\label{sec:quantitative}

In this section, we present a quantitative evaluation of exploit mitigation adoption (as per our baseline outlined in Section~\ref{backgroundsection}) and dependency support among popular embedded operating systems and hardware. The results of our quantitative evaluation reflect the current embedded \emph{state-of-the-art}, allowing us to identify clear gap-areas and new opportunities to improve the security of such systems.

\subsection{Embedded OS Mitigation and Dependency Support}\label{sec:embedded_os_mit_dep}

We evaluated 42 popular embedded operating systems to present an overview of the current state of embedded OS mitigation adoption. Our selection aims to be a representative sample of embedded operating systems and includes those listed by the UBM Embedded Markets Study~\cite{ubm_study}, and various studies into embedded operating systems~\cite{hahm_iotsurvey:2016,yerraballi_rtos:2000,elvstam_iot:2016} as well as some of the most popular mobile operating systems~\cite{netmarketshare_mobile_os:2017}.

We evaluated these 42 embedded OS for exploit mitigation and dependency
support through a combination of vendor surveys, documentation consultation,
and experimental validation. An overview of the results (including the list of
OSes) is shown in Tables~\ref{table:os_mitigation_adoption} and
\ref{table:os_feature_support}, while aggregated results are shown in Tables~\ref{table:mitigation_support_all} and \ref{table:feature_support_all}.

\begin{table}[!htbp]
	\tiny
	\centering
	\caption{Detailed Embedded OS exploit mitigation adoption. Red for Mobile embedded OSes, white for regular embedded OSes and gray for deeply embedded OSes.}
	\begin{tabular}{m{1.4cm}K{0.24cm}K{0.42cm}K{0.39cm}m{1.65cm}K{0.24cm}K{0.42cm}K{0.45cm}}
		\textbf{OS} & \textbf{ESP} & \textbf{ASLR} & \textbf{Canaries} & \textbf{OS} &\textbf{ESP} & \textbf{ASLR} & \textbf{Canaries}\\
		\hline	
		\rowcolor{LRed}	BlackBerry OS & \checkmark & \checkmark & \checkmark& Android\textsuperscript{$\ast$} & \checkmark & \checkmark & \checkmark \\
		\rowcolor{LRed} iOS\textsuperscript{$\ast$} & \checkmark & \checkmark & \checkmark & Win 10 Mob.\textsuperscript{$\ast$} & \checkmark & \checkmark & \checkmark \\
		\rowcolor{LRed} Sailfish OS\textsuperscript{$\ast$} & \checkmark & \checkmark & \checkmark & Tizen\textsuperscript{$\ast$} & \checkmark & \checkmark & \checkmark \\
		\rowcolor{white} Ubuntu Core\textsuperscript{$\ast$} & \checkmark & \checkmark & \checkmark &  Brillo\textsuperscript{$\ast$} & \checkmark & \checkmark & \checkmark \\
		\rowcolor{white}Yocto Linux\textsuperscript{$\ast$} & \checkmark & \checkmark & \checkmark & Windows Embedded\textsuperscript{$\ast$} & \checkmark & \checkmark & \checkmark \\
		\rowcolor{white} OpenWRT\textsuperscript{$\ast$} & \checkmark & \checkmark & \checkmark & Junos OS\textsuperscript{$\ast$} & \checkmark & $\times$ & \checkmark \\
		\rowcolor{white} $\mu$Clinux\textsuperscript{$\ast$} & \checkmark & $\times$ & \checkmark & CentOS\textsuperscript{$\ast$} & \checkmark & \checkmark & \checkmark \\
		\rowcolor{white} NetBSD\textsuperscript{$\ast$} & \checkmark & \checkmark & \checkmark & IntervalZero RTX\textsuperscript{$\ast$} & \checkmark & $\times$ & \checkmark \\
		\rowcolor{white} ScreenOS & $\times$ & $\times$ & $\times$ & Enea OSE & $\times$ & $\times$ & $\times$ \\
		\rowcolor{white} QNX & \checkmark & \checkmark & \checkmark & VxWorks & \checkmark & $\times$ & $\times$ \\
		\rowcolor{white} INTEGRITY & \checkmark & $\times$ & $\times$ & \redactedOS \textsuperscript{2} & $\times$ & $\times$ & $\times$ \\
		\rowcolor{gray!50} Cisco IOS & $\times$ & $\times$ & $\times$ & eCos & $\times$ & $\times$ & $\times$ \\
		\rowcolor{gray!50} Zephyr & \checkmark & $\times$ & \checkmark & ThreadX & $\times$ & $\times$ & $\times$ \\
		\rowcolor{gray!50} Nucleus & $\times$ & $\times$ & $\times$ & NXP MQX & $\times$ & $\times$ & $\times$ \\
		\rowcolor{gray!50} Kadak AMX & $\times$ & $\times$ & $\times$ & Keil RTX & $\times$ & $\times$ & $\times$ \\
		\rowcolor{gray!50} RTEMS & $\times$ & $\times$ & $\times$ & freeRTOS & $\times$ & $\times$ & $\times$ \\
		\rowcolor{gray!50} Micrium $\mu$C/OS\textsuperscript{1} & \checkmark & $\times$ & $\times$ & TI-RTOS & $\times$ & $\times$ & $\times$ \\
		\rowcolor{gray!50} DSP/BIOS & $\times$ & $\times$ & $\times$ & TinyOS & $\times$ & $\times$ & $\times$ \\
		\rowcolor{gray!50} LiteOS & $\times$ & $\times$ & $\times$ & RIOT & $\times$ & $\times$ & $\times$ \\
		\rowcolor{gray!50} ARM mbed & \checkmark & $\times$ & $\times$ & Contiki & $\times$ & $\times$ & $\times$ \\
		\rowcolor{gray!50} Nano-RK & $\times$ & $\times$ & $\times$ & Mantis & $\times$ & $\times$ & $\times$ \\
	\end{tabular}
	
	\label{table:os_mitigation_adoption}   
	\small \textsuperscript{ $\ast$} \emph{Embedded OS based on Windows, Linux or BSD.}\\
	\small \textsuperscript{1} \emph{A $\mu$C/OS-II kernel version with ESP support is available via a Micrium partner.}\\
	\small \textsuperscript{2} Due to the sensitive nature of \redactedOS, we have received it on the condition of anonymity for the vendor. \redactedOS is a real-time OS which is primarily being used for aerospace applications.

	\bigskip

	\tiny
	\centering	
	\caption{Detailed Embedded OS exploit mitigation dependency support for Memory Protection (MPROT), Virtual Memory (VMEM) and Random Number Generator (RNG). Red  for Mobile embedded OSes, white for regular embedded OSes and gray for deeply embedded OSes.}
	\begin{tabular}{m{1.4cm}K{0.5cm}K{0.4cm}K{0.3cm}m{1.4cm}K{0.6cm}K{0.5cm}K{0.4cm}}
		\textbf{OS} & \textbf{MPROT} & \textbf{VMEM} & \textbf{RNG} & \textbf{OS} & \textbf{MPROT} & \textbf{VMEM} & \textbf{RNG}\\
		\hline
		\rowcolor{LRed}	Android\textsuperscript{$\ast$} & \checkmark & \checkmark & \checkmark & iOS\textsuperscript{$\ast$} & \checkmark & \checkmark & \checkmark \\
		\rowcolor{LRed}	Win10 Mob.\textsuperscript{$\ast$} & \checkmark & \checkmark & \checkmark & BlackBerry OS & \checkmark & \checkmark & \checkmark \\
		\rowcolor{LRed}	Tizen\textsuperscript{$\ast$} & \checkmark & \checkmark & \checkmark & Sailfish OS\textsuperscript{$\ast$} & \checkmark & \checkmark & \checkmark \\
		\rowcolor{white} Ubuntu Core\textsuperscript{$\ast$} & \checkmark & \checkmark & \checkmark & Brillo\textsuperscript{$\ast$} & \checkmark & \checkmark & \checkmark \\
		\rowcolor{white} Yocto Linux\textsuperscript{$\ast$} & \checkmark & \checkmark & \checkmark & Windows Embedded\textsuperscript{$\ast$} & \checkmark & \checkmark & \checkmark\\
		\rowcolor{white} OpenWRT\textsuperscript{$\ast$} & \checkmark & \checkmark & \checkmark & Junos OS\textsuperscript{$\ast$} & \checkmark & \checkmark & \checkmark\\
		\rowcolor{white} $\mu$Clinux\textsuperscript{$\ast$} & \checkmark & \checkmark & \checkmark & CentOS\textsuperscript{$\ast$} & \checkmark & \checkmark & \checkmark \\
		\rowcolor{white} NetBSD\textsuperscript{$\ast$} & \checkmark & \checkmark & \checkmark & IntervalZero RTX\textsuperscript{$\ast$} & \checkmark & \checkmark & \checkmark\\
		\rowcolor{white} ScreenOS & \checkmark & \checkmark & \checkmark & Enea OSE & \checkmark & \checkmark & $\times$\\
		\rowcolor{white} QNX & \checkmark & \checkmark & \checkmark & VxWorks & \checkmark & \checkmark & $\times$\\
		\rowcolor{white} INTEGRITY & \checkmark & \checkmark & \checkmark & \redactedOS & \checkmark & \checkmark & \checkmark\\
		\rowcolor{gray!50} Cisco IOS & $\times$ & $\times$ & \checkmark & eCos & $\times$ & $\times$ & $\times$\\
		\rowcolor{gray!50} Zephyr& \checkmark & $\times$ & $\times$ & ThreadX & \checkmark & $\times$ & $\times$\\
		\rowcolor{gray!50} Nucleus & \checkmark & $\times$ & $\times$ & NXP MQX & \checkmark & $\times$ & $\times$\\
		\rowcolor{gray!50} Kadak AMX & $\times$ & $\times$ & $\times$ & Keil RTX & \checkmark & $\times$ & $\times$\\
		\rowcolor{gray!50} RTEMS & $\times$ & $\times$ & $\times$ & FreeRTOS & \checkmark & $\times$ & $\times$\\
		\rowcolor{gray!50} Micrium $\mu$C/OS & \checkmark & $\times$ & $\times$ & TI-RTOS & \checkmark & $\times$ & $\times$\\
		\rowcolor{gray!50} DSP/BIOS & \checkmark & $\times$ & $\times$ & TinyOS & \checkmark & $\times$ & $\times$\\
		\rowcolor{gray!50} LiteOS & \checkmark & $\times$ & $\times$ & RIOT & \checkmark & $\times$ & $\times$\\
		\rowcolor{gray!50} ARM mbed & \checkmark & $\times$ & $\times$ & Contiki & $\times$ & $\times$ & $\times$\\
		\rowcolor{gray!50} Nano-RK & $\times$ & $\times$ & $\times$ & Mantis & $\times$ & $\times$ & $\times$\\
	\end{tabular}
	
	\label{table:os_feature_support}   
	\small \textsuperscript{ $\ast$} \emph{Embedded OS based on Windows, Linux or BSD.}\\
	
\end{table}

Note that we consider a mitigation or feature supported \emph{\textbf{iff}} it is supported by the OS for at least some (but not necessarily all) platforms. Since this is a quantitative assessment, it neither evaluates the quality of the implementation nor whether the feature is enabled by default and as such the assessment is an optimistic one. We mark an OS as providing a CSPRNG (Cryptographically Secure Pseudo-Random Number Generator) \emph{\textbf{iff}} provided PRNG (Pseudo-Random Number Generator) functionality is advertised as such or can be reasonably assumed to provide secure random number generation functionality. 

\begin{table}[t]
\scriptsize
	\setlength\tabcolsep{1pt}
	\centering
	\caption {Overview of Embedded OS Exploit Mitigation Support.}
	\begin{tabular}{m{3.2cm}m{1.2cm}m{1.3cm}K{1.8cm}}
		\hline
		OS vs. Mitigation                 & ESP  & ASLR & Stack Canaries \\ \hline
		All evaluated OSes               & 20/42 & 13/42 & 17/42           \\ \hline
		Non-Mobile                   & 16/36 & 8/36 & 12/36           \\ \hline
		Non-Linux/Windows/BSD & 7/27 & 1/27 & 3/27           \\ \hline
		Deeply Embedded               & 3/20 & 0/20 & 1/20           \\ \hline
	\end{tabular}
	
	\label{table:mitigation_support_all}
\end{table}

\begin{table}[t]
\scriptsize
	\centering
	\caption {Overview of Embedded OS Exploit Mitigation Dependency Support.}	
	\begin{tabular}{m{2.4cm}m{2.0cm}m{1.8cm}m{1.34cm}}
		\hline
		OS vs. Mitigation                 & Memory Protection  & Virtual Memory  & OS CSPRNG  \\ \hline
		All OSes               & 34/42               & 21/42            & 20/42       \\ \hline
		Non-Mobile                  & 29/36               & 16/36            & 15/36       \\ \hline
		Non-Linux/Win/BSD & 20/27               & 5/27            & 6/27       \\ \hline
		Deeply Embedded               & 13/20               & 0/20            & 1/20       \\ \hline
	\end{tabular}
	
	\label{table:feature_support_all}
\end{table}

\subsection{Embedded Hardware Feature Support}\label{sec:embedded_hw_feat_sup}

\subsubsection{Von Neumann vs Harvard}

Before we can discuss the hardware features support for exploit mitigations in embedded systems, we first need to consider that there are essentially two main processor architectural styles: \emph{Harvard} and \emph{Von Neumann}.
The Harvard CPU architecture has separated instruction and data busses and thus allows operations to run simultaneously, while physically separating signals and storage for code and data memory. 
In contrast, the Von Neumann CPU architecture has only one bus which is used for both data transfers and instruction fetches, thus any value in memory can be executed or interpreted as data, respectively.

\subsubsection{Hardware Feature Support}\label{sec:embedded_hw_sup}

The embedded world features a wide range of different processor architectures
and \emph{core families} with different capabilities. To establish an overview
of common embedded hardware capabilities, we make a selection of several \emph{core families} and map out their architectural style and MPU, MMU, and hardware ESP support capabilities.

We evaluated 78 different \emph{core families} for hardware dependency support.  57 of core family SoCs were based on Von Neumann architecture and 21 of them were based on Harvard architecture. An overview of the supported feature detailed results reported in Tables~\ref{table:hardware_feature_support_a} and \ref{table:hardware_feature_support_b}. Our selection of core families aims to be a representative sample of \emph{core families} belonging to major architectures and vendors in the embedded space across industry verticals and includes, among others, the most popular \emph{core families} listed by recent UBM Embedded Markets Studies~\cite{ubm_study} and EDN reader surveys~\cite{mcu_popularity:2013}.

\begin{table}[!htbp]
\scriptsize
	\centering
	\caption{Core Family dependency support in Harvard (H) and Von Neumann (N) architectures. We consider a feature supported if it is supported by all members of a given core family and absent if it is not supported by any of them. Any variation with regards to dependency support is denoted with $\sim$ and omitted from aggregated results.}
	\begin{tabular}{lcccc}
		\textbf{Core Family} & \textbf{Arch.} & \textbf{MPU} & \textbf{MMU} & \textbf{ESP} \\
		\hline
		\textbf{ARM} & & & & \\
		\hdashline
		ARM1 & N & $\times$ & $\times$ & $\times$ \\
		ARM2 & N & $\times$ & $\times$ & $\times$ \\
		ARM3 & N & $\times$ & $\times$ & $\times$ \\
		ARM6 & N & $\times$ & $\times$ & $\times$ \\
		ARM7 & N & $\times$ & $\times$ & $\times$ \\
		ARM7T & N & $\sim$ & $\sim$ & $\times$ \\
		ARM7EJ & N & $\times$ & $\times$ & $\times$ \\
		ARM8 & N & $\times$ & \checkmark & $\times$ \\
		ARM9T & N & $\sim$ & $\sim$ & $\times$ \\
		ARM9E & N & $\sim$ & $\sim$ & $\times$ \\
		ARM10E & N & $\times$ & \checkmark & $\times$ \\
		ARM11 & N & $\sim$ & $\sim$ & \checkmark \\
		ARM Cortex-A & N & $\times$ & \checkmark & \checkmark \\
		ARM Cortex-R & N & \checkmark & $\times$ & \checkmark \\
		ARM Cortex-M & N & $\sim$ & $\times$ & \checkmark \\
		\hline
		\textbf{PIC} & & & & \\
		\hdashline
		PIC10 & H & $\times$ & $\times$ & $\times$ \\
		PIC12 & H & $\times$ & $\times$ & $\times$ \\
		PIC16 & H & $\times$ & $\times$ & $\times$ \\
		PIC18 & H & $\times$ & $\times$ & $\times$ \\
		PIC24 & H & $\times$ & $\times$ & $\times$ \\
		dsPIC & H & $\times$ & $\times$ & $\times$ \\
		\hline
		\textbf{MIPS32} & & & & \\
		\hdashline
		PIC32MX & N & $\times$ & $\times$ & $\times$ \\
		PIC32MZ EC & N & $\times$ & \checkmark & $\times$ \\
		PIC32MZ EF & N & $\times$ & \checkmark & \checkmark \\
		PIC32MM & N & $\times$ & \checkmark & \checkmark \\
		\hline
		\textbf{PowerPC} & & & & \\
		\hdashline
		PPC e200 & N & $\sim$ & $\sim$ & \checkmark \\
		PPC e300 & N & $\times$ & \checkmark & \checkmark \\
		PPC e500 & N & $\times$ & \checkmark & \checkmark \\
		PPC e600 & N & $\times$ & \checkmark & \checkmark \\
		PPC 403 & N & $\times$ & $\times$ & $\times$ \\
		PPC 401 & N & $\times$ & $\times$ & $\times$ \\
		PPC 405 & N & $\times$ & \checkmark & \checkmark \\
		PPC 440 & N & $\times$ & \checkmark & \checkmark \\
		PPC 740 & N & $\times$ & \checkmark & \checkmark \\
		PPC 750 & N & $\times$ & \checkmark & \checkmark \\
		PPC 603 & N & $\times$ & \checkmark & \checkmark \\
		PPC 604 & N & $\times$ & \checkmark & \checkmark \\
		PPC 7400 & N & $\times$ & \checkmark & \checkmark \\
	\end{tabular}
	\label{table:hardware_feature_support_a}
\end{table}

\begin{table}[!htbp]
\scriptsize
	\centering
	\caption{Core Family dependency support in Harvard (H) and Von Neumann (N) architectures II}
	\begin{tabular}{lcccc}
		\textbf{Core Family} & \textbf{Arch.} & \textbf{MPU} & \textbf{MMU} & \textbf{ESP} \\
		\hline
		\textbf{x86} & & & & \\
		\hdashline
		Intel Atom Z34XX & N & $\times$ & \checkmark & \checkmark \\
		Intel Quark X10XX & N & $\times$ & \checkmark & \checkmark \\
		Intel Quark $\mu$C\textsuperscript{1} & N & $\times$ & \checkmark & \checkmark \\
		\hline
		\textbf{SuperH} & & & & \\
		\hdashline
		SH-1 & N & $\times$ & $\times$ & $\times$ \\
		SH-2 & N & $\times$ & $\times$ & $\times$ \\
		SH-3 & N & $\times$ & \checkmark & $\times$ \\
		SH-4 & N & $\times$ & \checkmark & $\times$ \\
		\hline
		\textbf{AVR} & & & & \\
		\hdashline
		ATtiny & H & $\times$ & $\times$ & $\times$ \\
		ATmega & H & $\times$ & $\times$ & $\times$ \\
		ATxmega & H & $\times$ & $\times$ & $\times$ \\
		\hline
		\textbf{AVR32} & & & & \\
		\hdashline
		AVR32A & N & \checkmark & $\times$ & $\times$ \\
		AVR32B & N & $\times$ & \checkmark & $\times$ \\
		\hline
		\textbf{8051} & & & & \\
		\hdashline
		Intel MCS-51 & H & $\times$ & $\times$ & $\times$ \\
		Infineon XC88X-I & H & $\times$ & $\times$ & $\times$ \\
		Infineon XC88X-A & H & $\times$ & $\times$ & $\times$ \\
		\hline
		\textbf{m68k} & & & & \\
		\hdashline
		NXP M683XX & N & $\times$ & $\times$ & $\times$ \\
		NXP ColdFire V1 & N & $\times$ & $\times$ & $\times$ \\
		NXP ColdFire V2 & N & $\times$ & $\times$ & $\times$ \\
		NXP ColdFire V3 & N & $\times$ & $\times$ & $\times$ \\
		NXP ColdFire V4 & N & $\times$ & \checkmark & $\times$ \\
		NXP ColdFire V5 & N & $\times$ & \checkmark & $\times$ \\
		\hline
		\textbf{TriCore} & & & & \\
		\hdashline
		Infineon TC11xx & H & $\times$ & $\sim$ & $\times$ \\
		Infineon AUDO Future & H & $\times$ & $\times$ & $\times$ \\
		\hline
		\textbf{C166} & & & & \\
		\hdashline
		Infineon XE166 & N & \checkmark & $\times$ & \checkmark \\
		Infineon XC2200 & N & \checkmark & $\times$ & \checkmark \\
		\hline
		\textbf{MSP430} & & & & \\
		\hdashline
		MSP430x1xx & N & $\times$ & $\times$ & $\times$ \\
		MSP430x2xx & N & $\times$ & $\times$ & $\times$ \\
		MSP430x3xx & N & $\times$ & $\times$ & $\times$ \\
		MSP430x4xx & N & $\times$ & $\times$ & $\times$ \\
		MSP430x5xx & N & $\times$ & $\times$ & $\times$ \\
		MSP430x6xx & N & $\times$ & $\times$ & $\times$ \\
		MSP430FRxx & N & \checkmark & $\times$ & $\times$ \\
		\hline
		\textbf{Blackfin} & & & & \\
		\hdashline
		Analog Blackfin\textsuperscript{2} & N & \checkmark & $\times$ & $\times$ \\
		\hline
		\textbf{ARC} & & & & \\
		\hdashline
		Synopsys ARC EM & H & $\sim$ & $\times$ & $\times$ \\
		Synopsys ARC 600 & H & $\sim$ & $\times$ & $\times$ \\
		Synopsys ARC 700 & H& $\times$ & $\sim$ & $\times$ \\
		\hline
		\textbf{RL78} & & & & \\
		\hdashline
		Renesas RL78/G1x & H & $\times$ & $\times$ & $\times$ \\
		Renesas RL78/L1x & H & $\times$ & $\times$ & $\times$ \\
		\hline
		\textbf{RX} & & & & \\
		\hdashline
		Renesas RX200 & H & $\sim$ & $\times$ & $\times$ \\
		Renesas RX600 & H & $\sim$ & $\times$ & $\times$ \\		
	\end{tabular}

	\label{table:hardware_feature_support_b}
	\small \textsuperscript{1} \emph{Intel Quark Microcontrollers (D1000/C1000/D2000)}\\
	\small \textsuperscript{2} \emph{Although documentation mentions an MMU, it does not support address translation (and thus does not allow for virtual memory) which is why we consider it an MPU for our purposes.}
\end{table}

The \emph{core families} in our selection belong to the following embedded architectures: ARM, MIPS32, PIC, PPC, x86, SuperH, AVR, AVR32, Intel 8051, Motorola 68000, TriCore, MSP430, C166, Blackfin, ARC, Renesas Electronics RL78 and RX.

\begin{table}
\scriptsize
	\centering
	\caption{Overview of hardware dependency support in Von Neumann core families.}
	\begin{tabular}{m{4.8cm}K{2.5cm}}
		\hline
		Von Neumann Mitigation Support& Support\\ \hline
		MPU                                                     & 6/51 (11.8\%)  \\ \hline
		MMU                                                     & 24/51 (47.1\%)  \\ \hline
		Hardware ESP                                                     & 22/51 (43.1\%)     \\ \hline
	\end{tabular}
	\label{fig:core_stats}
\end{table}

\subsection{Quantitative Analysis Results}\label{sec:quantitative_conclusions}

Among the embedded OSes surveyed, we can distinguish two major clusters in terms of capabilities and purposes:

\begin{itemize}
	\item \textbf{High-End}: These OSes are aimed at the higher end of the embedded spectrum and offer \emph{virtual memory} capabilities as well as often being POSIX-compliant. This includes mobile OSes (e.g., Android and iOS), lightweight versions of OSes common in the general-purpose world (such as Linux, Windows or BSD) as well as OSes like QNX or VxWorks.
	
	\item \textbf{Low-End}: These OSes are aimed at \emph{deeply embedded systems}, often have real-time capabilities and do not offer \emph{virtual memory} support. As such, there is usually no separation between user- and kernelspace and instead of isolated processes, there tends to be just a kernel running a limited set of tasks. Examples are Real-Time Operating Systems (RTOSes) such as ThreadX, RTEMS, Micrium $\mu$C/OS and TinyOS.
\end{itemize}

From the results in Tables~\ref{table:mitigation_support_all} and \ref{table:feature_support_all}, we can observe that all mobile OSes have support for every exploit mitigation in our baseline and so do most Linux, BSD, and Windows-based OSes. Outside of those, however, almost all other OSes (apart from QNX) lack support for any mitigations whatsoever. We can also see that while memory protection support is almost universally present, virtual memory and OS CSPRNG support is almost universally lacking in the \emph{low-end} OSes aimed at deeply embedded systems. From these observations we can conclude that exploit mitigation adoption (and underlying dependency support) is generally present only on the \emph{high-end} embedded OSes which derive from Linux, BSD or Windows.

When it comes to the hardware \emph{core families} surveyed, we can see that less than half of the (Von Neumann) \emph{core families} in our selection have MMU support. A small minority of \emph{core families} has MPU support (MPU was supported in 6 out of 51 in Von Neumann architecture), leaving just under half of the Von Neumann \emph{core families} in our selection without necessary hardware support for memory protection and over half without the hardware support required for virtual memory. This lack of MPU and MMU support makes sense for the more constrained end of the spectrum such as MCUs, which only have support for integrated memory and no support for external memory. Similarly, under half of Von Neumann \emph{core families} in the selection have hardware ESP support, meaning ESP can only be implemented via software emulation on these systems. 

As observed by Barr~\cite{barr_embedded_software_trends:2012}, UBM Embedded Markets study~\cite{ubm_study} and other observers~\cite{clarke_mcu_32bit:2013}, the embedded world is seeing a trend towards deployment of 32-bit CPUs (and for the most high-end embedded systems even 64-bit CPUs~\cite{simon_embedded_64bit:2015}) over the traditionally used 8- or 16-bit CPUs. Since most popular 32-bit architectures are Von Neumann, this has security implications, though these are possibly offset by the fact that certain modern CPU architectures offer hardware ESP support (e.g., ARMv6+, MIPS32r3+, x86, etc.).

Based on the above observations, we can conclude that there is a gap when it comes to \emph{deeply embedded systems}. Only among the \emph{high-end} Linux, BSD and Windows-based OSes there are significant exploit mitigation adoption and when it comes to \emph{low-end} OS capabilities, the lack of virtual memory and cryptographically secure pseudorandom number generator (CSPRNG) support present obstacles to ASLR and stack canary adoption.

\section{Challenges for Embedded Systems}\label{challenges} %

\label{ch:challenges} %
To explain the gaps in embedded exploit mitigation adoption and implementation discussed in Section~\ref{sec:quantitative}, we now discuss the challenges faced by embedded systems developers for integrating mitigation within their software. Based on this discussion, we will identify a series of \emph{open problems} in the field of embedded exploit mitigations and outline the design criteria for exploit mitigations and OS CSPRNGs for \emph{deeply embedded systems}. 
Generally we can divide the reasoning behind lack of exploit mitigation within embedded system to the following groups:
\begin{enumerate}
	\item Development Practices \& Cost Sensitivity
	\item Resource Constraints
	\item Safety, Reliability \& Real-Time Requirements\
	\item Hardware \& OS Limitations
\end{enumerate}
We discuss these open problems to understand why a mitigation is not available even though the required hardware features exist or why hardware features are missing in practice.

\subsection{Development Practices \& Cost Sensitivity}\label{sec:devpract_cost}

Embedded systems development practices and design cultures are different~\cite{koopman1996embedded} from those in desktop or web application development. Compared to the general-purpose world, the embedded world is heavily fragmented~\cite{van2002integrating} among many different vendors and suppliers and technologies themselves are fragmented into competing standards without clear market leaders and individual solutions for specific problems. For example, a typical embedded product is put together as the result of a hardware vendor selling a chip, deployed with an operating system and some drivers, to an embedded systems manufacturer (the \emph{Original Equipment Manufacturer (OEM)}) who integrates it into the embedded system in question (adding some hardware peripherals, writing some software) and often resells it to a brand-name company who adds a user- or machine-to-machine interface and put it on the consumer market.  This fragmentation leads to the following issues:

\begin{itemize}
	
\item \textbf{{Lowest Common Denominator}} Vendors at the top of the chain such as chip or embedded operating system vendors often cater to very diverse customers and as such are bound by the demands (in terms of capabilities, overhead and cost increases) of their most constrained customers.
\item \textbf{{Fragmented Security Requirements}} No single entity oversees the entire software development life-cycle and as such, there is no coherent, single set of security requirements.
	
\item  \textbf {{Patching \& Maintenance Issues}} In many cases no single entity has the ability to patch or upgrade every piece of software on a given embedded device once it's shipped.
\item \textbf{{Incentive Issues}} 
	While there might be a strong case for certain security measures when considering the end product as a whole, there is usually little incentive on part of individual vendors and manufacturers who are just a single link in a much bigger chain. 
	Embedded systems markets are often characterized~\cite{solingen_embedded_softdev:2002,koopman_embedded_design_issues:1996} by a heavy focus on time-to-market (earlier market introduction tends to mean deeper market penetration and hence higher potential revenue) and novel features: since embedded systems are designed for a specific purpose rather than general-purpose computing, vendors often differentiate themselves on the basis of price and specific features rather than generic capabilities unrelated to the specific utility of the end product. As a result, there is little incentive for integrating security measures if these are not already present by default.
\item \textbf{{Cost Sensitivity}} Embedded systems are often very cost sensitive~\cite{koopman_embedded_system_security:2004}, they tend to be produced in large quantities and as such even small cost increases per unit rapidly amount to large overall production cost increases. In addition, for the cheaper products a cost increase on part of a single component soon amounts to a higher percentage of total system cost, making it harder to justify such a cost increase, especially with something that is often so hard to quantify as \emph{improved security}.

Finally, any cost savings might aid in gaining a market advantage for price sensitive products. As such there usually is a preference for cheaper, simpler hardware such as chips with few features and limited room for overhead. 
This matters from a security perspective because it makes many hardware-based security measures infeasible (because of the associated per-unit cost increases) and means designers and implementers of embedded security measures have to deal with limited hardware capabilities and resource constraints.

\end{itemize}

\subsection{Resource Constraints}\label{sec:resource_constraints}

Generally, embedded systems face significant resource constraints~\cite{koopman_embedded_design_issues:1996} since they are designed with a specialized, dedicated purpose in mind rather than aiming to provide general-purpose capabilities. As such, all resources considered superfluous to this task are eliminated to reduce production cost which results in limitations on code and data memory, processing power and hardware capabilities. Embedded software, in turn, is designed to be efficient and have a minimal footprint in order to meet these constraints given the limited room for overhead. These constrains are including code storage size, memory size, processing power, and power consumption. An example of impact of mentioned constrains on exploit mitigation is power consumption. This constraint immediately conflicts with security measures that introduce power consumption overhead (e.g., due to being computationally intensive or requiring ``power hungry'' hardware).

In the context of this work we concern ourselves with four major resource constraint areas:

\begin{enumerate}
	
	\item \textbf{{Code Storage Size}}: This constraints limit code size overhead and the introduction of additional functionality. Many embedded systems are diskless and do not have permanent storage, storing code in flash memory of a few KB or MB instead. Those systems that do have permanent storage use something like a few KB of EEPROM, usually to store configuration data only since the infrequent changing of code means it is more economical to arrange this via flashing a firmware update, or are far more limited than hard disk capacities of desktop or server systems (e.g., using SD cards of a few GB).
	
	\item \textbf{{Memory Size}}: This  constraints limit memory usage overhead and often rule out the possibility of memory-intensive computations. Embedded systems, particularly \emph{deeply embedded systems}, often do not have external memory but rely only on a few KB or MB of on-chip internal (S)RAM. Those systems that do have external memory are often limited to anything from a few dozen MB up to one or two GB.
	
	\item \textbf{{Processing Power}}: While there is a trend towards usage of more powerful 32-bit processors~\cite{barr_embedded_software_trends:2012,ubm_study,clarke_mcu_32bit:2013} running at clock speeds ranging from 100 MHz to around 1 GHz (the average in 2015 being 397 MHz according to ~\cite{ubm_study}) and there are plenty of embedded segments where even more serious computing power is a must, many embedded systems continue to use simpler 8- or 16-bit processors with clock speeds ranging from 8 to 32 MHz. Such a lack of processing power inhibits deployment of computationally intensive security measures and certain cryptographic algorithms.
	
    \item \textbf{{Power Consumption}}: Many embedded systems have serious power consumption constraints~\cite{koopman_embedded_design_issues:1996,koopman_embedded_system_security:2004} as a result of being battery operated, having to last months, years or indefinitely on a single battery while others might get recharged more frequently. As such, this constraint conflicts with security measures that introduce significant power consumption overhead (as mentioned above).
	
\end{enumerate}

\subsection{Safety, Reliability \& Real-Time Requirements}

Embedded systems tend to have specific requirements relating to safety, reliability, and real-time computation~\cite{koopman_deep_survivability:2007}:

\paragraph{Safety \& Reliability} Some embedded systems have stringent safety and reliability requirements which would require certification of any security measures upon their introduction and require them to be robustly reliable (e.g., maintain availability). This means, for example, that exploit mitigations for these embedded systems will have to avoid invocation of \emph{alert policies} that violate safety and reliability (such as abruptly terminating critical software upon detection of attacks~\cite{Abbasi:2017:EAC:3134600.3134618,abbasi2018race}).
\paragraph{Timeliness} Many embedded systems are subject to varying degrees of \emph{hardness} real-time requirements and use \emph{real-time operating systems (RTOS)} to accommodate this. As such, security measures for those systems will need to respect those requirements~\cite{Abbasi:2017:EAC:3134600.3134618}. However, such requirements might inherently conflict with certain exploit mitigation designs or their dependencies. Consider, for example, \emph{ASLR} and its dependency on \emph{virtual memory}. Traditionally, the use of \emph{virtual memory} in real-time operating systems has been avoided due to timing analysis complications~\cite{puat_rt_paging:2007}. \emph{Virtual memory} poses predictability problems regarding \emph{worst-case execution time (WCET)} analysis largely because of two issues~\cite{puat_rt_paging:2007,puffitsch_predictable_vm:2016}:
	
	\begin{enumerate}
		
		\item \textbf{Address Translation}: Mapping virtual to physical addresses is commonly done using a \emph{translation look-aside buffer (TLB)}: a memory cache that is part of the MMU and stores recent address translations. However, address translation timings are hard to predict, because a) not all mappings are cached in the TLB leading to cache misses requiring a subsequent page table lookup and b) the TLB is shared between different processes.
		
		\item \textbf{Paging}: Since physical memory is shared between different processes and any physical page may be selected for replacement by the paging algorithm, predicting whether a virtual memory reference results in a page fault is hard. In addition, paging makes memory access timings dependent on TLB and cache contents increasing unpredictability. Finally, page faults may incur significant overhead rendering a system non-responsive for too long.
		
	\end{enumerate}
	Various hardware/software-based proposals for real-time compatible \emph{virtual memory} exist~\cite{puat_rt_paging:2007,puffitsch_predictable_vm:2016}, but to the best of our knowledge, none of these have seen adoption by popular RTOSs due to significant performance penalties or hardware cost increases.

\subsection{Hardware \& OS Limitations}\label{sec:hw_and_os_limitations}

As a result of the embedded cost sensitivity and resource constraints discussed above, embedded hardware and operating systems are often lacking the features upon which modern security measures depend. We will briefly discuss the implications of these limitations for the future embedded adoption of the exploit mitigations in our baseline as well as identify some related open problems. 

\subsubsection{MPUs, MMUs \& Hardware ESP}

As shown in Table~\ref{fig:core_stats}, under half of surveyed embedded core families have \emph{hardware ESP} support. While 32-bit processors are clearly gaining increasing traction within the embedded world and are even displacing 8- and 16-bit processors~\cite{barr_embedded_software_trends:2012}, smaller 8-bit processors continue to dominate a significant portion of the embedded space. While many modern 32-bit processors tend to be Von Neumann and while many popular architectures in this category have hardware ESP support (e.g., ARMv6+, MIPS32r3+) there are others which do not.
Even though for many systems based on those smaller 8- and 16-bit processors it's quite reasonable to migrate to popular Harvard architectures (e.g., AVR, 8051, PIC, etc.), many modern 32-bit processors tend to be Von Neumann and while many popular architectures in this category have hardware ESP support (e.g., ARMv6+, MIPS32r3+) there are others which do not have such support. Considering that older Von Neumann processors will continue to be produced and integrated into new systems, this will leave a segment of embedded devices without hardware ESP support which is an open problem. 
Additionally, existing software ESP solutions (e.g., PaX's \emph{NOEXEC}) only support a limited number of OS and architecture combinations. As such, low-overhead software ESP support for a wide range of common embedded operating systems and processor architectures is currently an open problem.
Additionally, while Table~\ref{table:feature_support_all} shows that the majority of embedded operating systems offer memory protection support, not all embedded hardware offers the required underlying features to allow the OS to make use of this support. Table~\ref{fig:core_stats} shows only 47\% of all surveyed core families have MMU support and only 12\% have MPU support, which leaves 41\% unable to accommodate memory protection. Due to cost sensitivity as discussed in Section \ref{sec:devpract_cost}, embedded systems manufacturers are unlikely to migrate to costlier higher-end processors with MPU/MMU support mainly for security reasons and as such this leaves us with the open problem.
\subsubsection{Virtual Memory}

As discussed earlier, real-time requirements and lack of MMU support adversely affect embedded virtual memory adoption. While Table~\ref{table:feature_support_all} shows that 50\% and 44\% of all analyzed systems and all \emph{non-mobile} embedded operating systems offer virtual memory support, this drops to a mere 19\% if we eliminate the Linux-, BSD- and Windows-based ones. When it comes to \emph{deeply embedded systems}, virtual memory support is absent altogether. One also needs to take into account that even if an embedded OS offers virtual memory support, disk-less embedded systems cannot use this to extend RAM since this would requiring swapping to disk. All these constraints are so intrinsically tied to the embedded space that it is highly unlikely that we will see universal virtual memory adoption and as such the lack of alternatives to ASLR suitable for embedded systems without virtual memory remains an open problem.
\subsubsection{Advanced Processor Features}

Many modern security measure proposals rely on advanced processor features to offset otherwise unacceptable overhead penalties. Such features range from support for trusted computing (e.g., \emph{ARM TrustZone}), complication of kernel-mode exploitation, isolation of code and data regions in memory and pointer bounds checking to features utilized to support \emph{Control-Flow Integrity (CFI)} as well as cryptographic hardware acceleration.

When it comes to embedded systems, the problem with security measures which rely on such advanced processor features is that they are only available on the newest and most high-end architectures. Even among the more high-end embedded-oriented processors such as the \emph{Intel Atom} or the ARMv8-based CPUs the vast majority of these features is unsupported. 
Additionally, such advanced processor features are not likely to be adopted by any embedded-oriented processors other than the most high-end ones anytime soon either, considering the corresponding cost increase. 
As such, any proposal for embedded security measures seeking widespread adoption will need to avoid relying on such advanced processor features.
\subsubsection{OS CSPRNGs}

Secure randomness plays a fundamental role in the wider security ecosystem, not only for cryptographic purposes but also as a dependency upon which exploit mitigations rely. 
Since the design and implementation of a CSPRNG is not a trivial affair, the provision of secure randomness can be considered an important OS service. But as can be seen in Table \ref{table:feature_support_all}, OS CSPRNG support is far from universal in embedded operating systems. This is particularly visible in the non Linux-, BSD- and Windows-based operating systems and even more so in those aimed at \emph{deeply embedded systems}. 
Porting existing OS CSPRNG designs from the general-purpose world to the embedded world, even if it is from a GP-oriented version to an embedded-oriented version of the same operating system, is far from trivial for various reasons which we describe in the following.

\paragraph{OS \& Hardware Diversity} As discussed earlier in this work, the embedded world is heavily fragmented. The fact that embedded operating systems often seek to cater to platforms with much more divergent capabilities than their general-purpose counterparts means it is hard to identify universally available, suitable entropy sources. So while there exists a sizeable body of work around the design of embedded random number generators, these designs are generally very domain-specific as they rely on entropy sources (e.g., sensor values~\cite{corin_tinykey:2011}, radio and GPS data~\cite{francillon_tinyrng:2007}) present only in specific embedded devices.
\paragraph{Resource Constraints} The resource constraints discussed in Section \ref{sec:resource_constraints} also impact embedded PRNG design. Limited processing power, memory and code size constraints translate to a need for \emph{lightweight cryptography}~\cite{nist_lightweight_crypto:2017}: small, fast algorithms which still offer the appropriate degree of security. In addition, power consumption constraints necessitate a PRNG design that avoids constant entropy collection, especially considering many battery-operated \emph{deeply embedded} devices spend most of their time in standby modes waiting for event- or time-based activation to preserve battery life.
\paragraph{Low Entropy Environment} Perhaps the biggest hurdle in embedded PRNG design is the fact that embedded systems are generally a low entropy environment. Since they are designed for specific, limited tasks. 
	 In the general purpose world, where one can assume most systems have user peripherals and disks one can use the associated system events (e.g., keystroke timings, mouse movements) as a source of entropy. But for most embedded systems, being headless and/or disk-less as well as having no user interaction, this is not an option.

	Ideally, this problem would be solved by having omnipresent, on-chip \emph{True Random Number Generator (TRNG)}  available but considering embedded cost sensitivity issues this is not realistic. So in practice, one sees a lot of workarounds of dubious quality, which tend to lead to security issues of their own. Common and insecure approaches are to use \emph{personalization data} (e.g., device MAC addresses or serial numbers) as seed entropy~\cite{lorente_scrutinizing_wpa2:2015} or rely on manufacturer-supplied initial entropy, sometimes in the form of a so-called \emph{seed file}. But care needs to be taken here that these seed files are unique per device, unpredictable and secret. 
	This approach still leaves various problems for embedded systems such as dealing with disk-less nodes and not being applicable to the first system boot (which is often when embedded devices generated their long-term cryptographic keys).

\section{Summary of Open Problems}\label{sec:challenges_open_problems}

Based on the mentioned constraints and our quantitative results, we can identify two pressing open problems relating to embedded exploit mitigation adoption namely: \textbf{exploit mitigation} and \textbf{OS CSPRNG design} for deeply embedded systems.

\subsection{Deeply Embedded Exploit Mitigation Criteria}\label{subsec:embedded_exploit_mitigation_criteria}
We can distill the following criteria for \emph{deeply embedded} exploit mitigations based on the observations in Section~\ref{ch:challenges}:
\begin{enumerate}
	\item \textbf{Limited Resource Pressure}: Mitigations should limit pressure on constrained resources to a minimum and provide low \emph{worst-case} (rather than \emph{average-case}) overhead upper bounds. As observed by Szekeres et al.~\cite{eternalwar}, exploit mitigations are only likely to see widespread industry adoption if the \emph{average-case} imposed code size, memory and runtime performance overhead is between at most 5 and 10\%. %
	\item \textbf{Hardware Agnostic}: Mitigation designs should be hardware agnostic to widen deployability across the embedded hardware. This rules out any dedicated hardware proposals and any reliance on specific hardware features that are not commonly available in deeply embedded systems. This does not include hardware features commonly but not universally available such as hardware ESP.
	
	\item \textbf{Availability Preservation}: Mitigations should offer multiple measures to take upon detection of an attack that allows for different degrees of availability preservation, ranging from those that allow an attack to take place without interfering to those that reduce availability disruption to a minimum. The rationale behind the former is that if availability is of prime importance, the worst-case scenario for an exploited vulnerability is to disrupt this availability and as such an unhindered but reported attack that gains control of the system and keeps it up is preferable over a prevented attack that brings it down in the process.
\item \textbf{Real-Time Friendly}: Mitigations should not violate real-time requirements and as such avoid non-deterministic constructs. As discussed earlier, this rules out designs relying on \emph{virtual memory}. 
\item \textbf{Easy (RT)OS Integration}: Mitigations should be easy to integrate into existing (RT)OS without requiring significant redesign of the operating system itself to widen deployability across the embedded OS and reduce integration cost. 
\end{enumerate}

\subsection{OS CSPRNG Design for Deeply Embedded Systems}\label{subsec:non_domain_embedded_os_csprng_criteria}
We can distill the following criteria for \emph{non-domain specific deeply embedded OS CSPRNGs} based on the observations we made in Section~\ref{ch:challenges}:
\begin{enumerate}
	\item \textbf{Lightweight Cryptography}: The CSPRNG will have to be based on lightweight cryptographic primitives~\cite{nist_lightweight_crypto:2017} to accommodate code \& data memory as well as processing power constraints. Any OS CSPRNG design targeting \emph{deeply embedded systems} should be deployable on a representative hardware platform and only utilize a small fraction of available resources.
	\item \textbf{Entropy Gathering Limitations}: The CSPRNG will have to be designed in such a way as to not rely on constant runtime entropy gathering to reduce power consumption. This means entropy collection will have to be rapid and preferably take place mostly during system startup, given that many battery-operated embedded systems are in standby or powered off between small periods of event- or time-triggered activity.
\item \textbf{Non-Domain Specific Entropy Sources}: The CSPRNG will have to draw upon entropy sources that are both suitable in terms of entropic quality as well as nearly universally present in \emph{deeply embedded systems}. Ideally, such entropy sources have high throughputs so sufficient entropy is rapidly available at system startup and runtime entropy gathering can be limited. While there is nothing preventing CSPRNG augmentation with additional platform- or device-specific entropy sources (e.g., sensor values), these should not be the primary sources nor should the choice of entropy sources be left up to the system integrator.
\end{enumerate}

%% file: design.tex
\section{$\mu$Armor Design}\label{designsection}

In this section, we propose $\mu$Armor, an exploit mitigation
and OS CSPRNG baseline design for \emph{deeply embedded systems} in the form of LLVM passes. $\mu$Armor seeks to address the relevant gap areas based on the results from our quantitative analysis.  $\mu$Armor is targeted at those deeply embedded systems which satisfy the following conditions:

\begin{itemize}
	
	\item Feature either a (modified) Harvard architecture CPU or Von Neumann one with an MPU with hardware ESP support.
	
	\item Run a \emph{low-end} deeply embedded OS (e.g. Zephyr, FreeRTOS, TinyOS) or kernel with a single address space and without \emph{virtual memory} support. The OS is allowed to be multiple-stack, multi-threading, and real-time capable. 
\end{itemize}

\subsection{Attacker Model}

We assume an attacker who is capable of exploiting \emph{memory corruption vulnerabilities} within the target deeply embedded OS. Attacker wants to use such vulnerabilities to execute arbitrary code or invoke arbitrary system functionality. We assume that the attacker attempts to exploit a vulnerability over a networking protocol (e.g., Ethernet or WiFi). We assume the attacker does not have access to the specific firmware image of the target device, but she may have access to the firmware image of another instance of the system. Finally, $\mu$Armor does not seek to protect against \emph{data-flow hijacking} or \emph{data-only} attacks.

\subsection{High-Level Design}

\uArmor incorporates three mitigations measures in order to match the functionality of the baseline outlined in Section~\ref{backgroundsection}: \uESP, \uScramble, and \uSSP in the form of LLVM passes. In addition, it includes \uRNG in order to provide required OS CSPRNG support.

\subsection{$\mu$ESP Design}\label{sec:design_uesp}

\uESP is the \emph{Executable Space Protection (ESP)} component of \uArmor and unlike other ESP implementations is explicitly designed for MCUs running single address space OSes. \uESP assumes the OS can be modified for ESP-compliance, i.e., it allows for separation of code and data memory regions as well as avoiding code constructs such as dynamically generated code, stack-stored trampolines, etc. \uESP explicitly sets the hardware ESP non-executable bit for every memory region belonging to a non-code region, while not setting it for those owned by code region. Also, it ensures that no write permissions are set for memory regions belonging to a code region to avoid code modification attacks.

Furthermore, dynamic data memory objects (e.g., stack and heap) are ensured to be \emph{fully} placed in a data memory region, and the stack is instructed to grow \emph{away} from other data regions.  
Since most stacks grow downward, by placing the stack at the bottom of data memory, an overflow will cause an exception either because it is a code region in RAM (thus caught by \uESP permissions) or because we try to write outside of RAM. In a multi-stack environment, individual stacks overflowing into each other could be captured by placing a \emph{guard} at the end of each stack if MPU granularity and region count allows for this.

Finally, after setting up permissions, \uESP will mark any code regions responsible for MPU interaction or flash rewriting (e.g., in the bootloader) as non-executable. The reason behind such change is to avoid code-reuse attacks targeting these regions for permission-changing payloads or \emph{ret2bootloader} attacks~\cite{francillonharvardinjection,goodspeed_sage_advice:2013,zeronightsavr} and will disable further changes to the MPU by making the relevant control registers non-writable.

\subsection{$\mu$Scramble Design}\label{sec:design_uscramble}

$\mu$Scramble is a compile-time code diversification scheme which imposes minimal runtime overhead. Compile-time diversification allows us to leverage the high-level information available to the compiler, target multiple hardware platforms implicitly, avoid the need for disassembly and binary analysis as well as operate in an automatic fashion (as part of the regular compilation process) transparent to software developers. 

Note that the choice between diversification at compile-time requires infrastructures from the vendor. However, we believe that making \uScramble as a compile-time code diversification is reasonable due to results discussed on Section~\ref{ch:challenges} and major resource constraints exist within deeply embedded systems. 

Since the goal of \uScramble is to thwart code-reuse attacks, we aim to either \emph{eliminate} gadgets, \emph{randomize} them or make it \emph{infeasible to guess} their location in memory. \uScramble seeks to achieve this by diversifying code in a fine-grained manner at compile-time. Additionally, since \uScramble needs to be \emph{semantics-preserving} and to take into account embedded resource constraints regarding code size, memory usage and performance overhead we have created the diversification transformations listed below for \uScramble:

\paragraph{Register-Preservation Reordering} Most architectural calling conventions specify which registers are callee-saved and which are considered \emph{scratch} registers. Compilers take note of all registers used within a given subroutine and will ensure that those which are callee-saved are stored to the stack during the function prologue and restored from it during the epilogue. Such sequences are one of the most common targets for code-reuse gadgets due to their ability to act as register-setters terminated by a return instruction. Since the exact \emph{order} in which these registers are saved to and restored from the stack does not matter, we can randomize it and thus break gadget chain assumptions about what values end up in what registers.

\paragraph{Dead Code Insertion}
\uScramble supports two types of dead codes namely, \emph{no-operation (NOP)}-equivalent instructions that do not present opportunities for \emph{unaligned instruction gadgets}  and  \emph{trap instructions} which activate a violation policy handler upon execution (something which never happens during regular execution). The latter has the benefit of raising alerts while any attempt to brute-force gadgets is \emph{in progress}. 
	
\paragraph{Function Reordering}
	Unlike common function reordering techniques~\cite{giuffrida2012enhanced}, here due to embedded systems limitation, we randomize only the function order and as such the degree of diversification introduced is determined by the number of functions present in the target code.

The above transformations affect both code topology and code itself by randomizing the offsets of a) instructions with respect to a function address, b) functions with respect to the image base and c) one function with respect to another function as well as randomizing the order of register preservation code. Due to the fine-grained nature of our diversification we reduce memory object \emph{correlation} and offer better protection against information leaks and brute-force attacks than coarse-grained schemes.

\subsection{$\mu$SSP Design}\label{sec:design_ussp}

\uSSP is a component of \uArmor which ensures proper separation of \emph{data} and \emph{pointers} within a given local stackframe. \uSSP  works by placing the latter below the former so that stack overflows cannot target code or data pointers residing on the stack while the stack canary shields the stackframe metadata (e.g., saved frame pointer, return address). 
\uSSP also complies with regular GCC Stack Smashing Protection (SSP) function coverage parameters and is capable of protecting all kernel- and application-code that runs after early kernel and C support initialization.

On the operating system side, \uSSP uses a single master canary generated once at system boot for all OS tasks and threads. Since on deeply embedded systems without virtual memory there is no memory isolation for OS tasks nor a separation between kernel- and userspace, periodic canary renewal would lead to synchronization conflicts for a shared canary. Our solution for this problem would be assigning a dedicated master canary for each thread (or possibly only OS tasks) as well as one for the kernel and renewing canaries upon thread startup. The problem here, however, is that without virtual memory different threads utilize shared code which would require the compiler to figure out which code is used exclusively by a given thread and which code is shared, assigning a single common master canary for all shared code to prevent synchronization problems. This limits renewal effectiveness to such a degree, especially compared to incurred overhead cost, that we simply opt for the single master canary approach. In addition to that, the single address space nature (and accompanying lack of privilege separation) of most deeply embedded OSes would render a multi-canary scheme rather irrelevant as well.

As far as canary generation is concerned, \uSSP assumes the presence of either an OS CSPRNG (e.g., \uRNG described in Section~\ref{sec:design_urng}) or a TRNG (True Random Number Generator). We provide OS CSPRNG support as part of \uRNG.

The final components of \uSSP is its modular canary violation handlers. Deeply embedded exploit mitigations should offer multiple courses of action to be taken upon attack detection to allow for different degrees of availability preservation. We provide (in case the OS do not support it) different type of handlers which are discussed in Section~\ref{sec:sspimplement}.

\subsection{$\mu$RNG Design}\label{sec:design_urng}
\uRNG is our modification of a compact, software-only CSPRNG design for ARM Cortex-M by Van Herrewege et al.~\cite{van2014software} with a 128 bit security strength level. It utilizes the lightweight \texttt{Keccak}~\cite{bertoni2011keccak} sponge function as a CSPRNG~\cite{bertoni}. 
For obvious reasons, we do not create our own CSPRNG function. However, \uRNG is not a simple reimplementation of Keccak~\cite{van2014software}. \uRNG is extended to function as an OS CSPRNG by adding reseed control suitable to constrained embedded systems. More specifically, it avoids the constant runtime polling for reseeding, typical in most OS CSPRNG designs which puts a strain on power consumption. 
The main purpose of \uRNG in the context of this work is to serve as a dependency for exploit mitigations (such as stack canary mechanisms) but depending on chosen security strength, it is perfectly suitable as a general-purpose CSPRNG. While, for the sake of convenience, this work describes and implements \uRNG in the context of our representative platform, the \uRNG design is not restricted to any OS or platform in particular.

In \uRNG entropy accumulation is done by Keccak sponge \emph{absorption} functionality while random number generation is done by \emph{squeezing} the \texttt{Keccak} sponge, allowing us to use the same algorithm for both purposes. \uRNG uses the \texttt{Keccak-f[200]} permutation with \emph{rate} and \emph{capacity} parameters \texttt{r = 64} and \texttt{c = 136} respectively. Note that \uRNG is fully reseeded every $1 GB$ of output which results in a \texttt{Keccak} internal state of 25 bytes and generation of 64-bit pseudo-random numbers per \emph{squeeze} operation. We initially seed \uRNG with at least \texttt{256} bits of entropy and ensure reseeding is done with at least \texttt{256} bits of entropy.

When designing reseed control we need to take into account the applicability of \emph{passive} and \emph{active} state recovery attacks~\cite{bertoni}. In case of the former, the attacker cannot influence seed data while in case of the latter the attacker can. As per the original design by Van Herrewege et al.~\cite{van2014software} upon which \uRNG is based, \uRNG provides the required security against \emph{passive} state recovery attacks as long as reseeding occurs at least every $r * 2^{\frac{r}{2}} = 64 * 2^{32} = 32 GB$ of PRNG output and against \emph{active} attacks as long as reseeding happens at least every $2^{8} * r = 256 * 64 = 2 KB$ of output. Since our attacker model explicitly assumes a \emph{remote} attacker, incapable of influencing our entropy sources remotely, we only take \emph{passive} state recovery attacks into account. We have to consider a trade-off between overhead and security with respect to reseed frequency: ideally reseeding is done regularly to keep as much entropy in the PRNG as possible at all times but frequent entropy gathering puts pressure on embedded resources in terms of memory and power consumption.  \uRNG has two options for reseed control:

\begin{itemize}
	
	\item \textbf{Consistent}: 
	Reseed control here is integrated into the PRNG output function, ensuring at least 1 bit of entropy is accumulated for every 64 bits of PRNG output, thus ensuring a full 256-bit reseed every $2 KB$ of output.
	
	\item \textbf{Periodic}: 
	 Reseed control is integrated into the PRNG output function as well, together with a 32-bit reseed counter which keeps track of the number of bytes output, but actual reseed functionality is only invoked after the counter exceeds a certain threshold value \texttt{T}. Reseed functionality is designed to run for at most \texttt{S} seconds (to facilitate worst-case timing estimates) and accumulate entropy while resetting the reseed counter. 

\end{itemize}

Additionally, \uRNG uses non-domain specific entropy sources that can be found on most embedded devices. We can divide these sources into two groups:

\paragraph{\textbf{Initialization}} Initial entropy is gathered during early boot and should be rapidly available in sufficient quantity upon system startup to avoid the so-called \emph{boot-time entropy hole}~\cite{weakkeys}. We follow ~\cite{van2014software} in using \emph{SRAM Startup values (SUVs)} as our primary source of initial entropy. Using SRAM SUVs as a source of initial entropy allows us to have an entropy source that is present on most embedded devices, instantly available in (very) early boot and differs from boot session to boot session as well as from device to device. 
		As discussed in the literature~\cite{pufavr,pufthesis,van2014software}, the amount of entropy in modern microcontroller SRAM tends to be around 5\% of its total size at normal operating temperatures. This means that, on average, \uRNG would require at least $\frac{2 * 128}{0.05 * 8} = 640$ bytes of SRAM to guarantee a security strength level of 128 bits, a reasonable restriction for most modern microcontrollers.

\paragraph{\textbf{Reseeding}} Reseed entropy is gathered upon invocation of reseed control functionality and should provide either at least 1 bit of entropy per invocation (in case of \emph{consistent} reseed control) or an appropriate throughput rate (in case of \emph{periodic} reseed control).
	
	\begin{itemize}
		
		\item \textbf{Clock Jitter \& Drift}: The various oscillators (e.g., RC, Ring or VC oscillators) acting as microcontroller clock signal sources are never completely stable and are influenced by factors such as supply voltage, temperature, etc.

		\item \textbf{ADC Noise}: Many embedded systems are outfitted with \emph{analog-to-digital converters (ADCs)} we can use it as an entropy source by sampling the least significant bit of ADC output corresponding to floating inputs. It is worth mentioning that general suitability of ADC Noise as cryptographic entropy source is unevaluated and as such we recommend against integration unless proper device-specific evaluation indicates suitability.
		
	\end{itemize}

%% file: implementation.tex
\section{$\mu$Armor Implementation} \label{ch:implementation}

In the following, we describe implementation aspects for all $\mu$Armor components.

\subsection{Representative Platform}\label{sec:representative_platform}

$\mu$Armor is not intrinsically tied to any particular OS or hardware configuration. However, for our implementation we choose \emph{Zephyr}~\cite{zephyr:2017} RTOS on a \texttt{TI LM3S6965} microcontroller, which is based on a 50 MHz ARM Cortex-M3 outfitted with 256 kB flash, 64 kB SRAM and an MPU. We chose \emph{Zephyr} because it is an actively developed, open-source OS with a permissive license aimed at resource-constrained embedded devices and is supported by the \emph{Linux Foundation} and major chip vendors such as \emph{Intel}, \emph{NXP}, and \emph{Synopsys}.
We picked the \texttt{TI LM3S6965} microcontroller because it is supported by \emph{Zephyr} and is representative of a typical \emph{deeply embedded system} with limited resources. As noted above, the $\mu$Armor protection scheme is implemented as LLVM passes.

\subsection{$\mu$ESP Implementation}
We consider the following two common approaches to deal with code and data memory in embedded systems: (i) program code is located in and executed from flash and data is copied to RAM and (ii) both program code and data are copied from flash to RAM by a first stage bootloader and further code (e.g., the OS kernel) is executed from RAM. For \uESP this corresponds to the permission policies outlined in Table~\ref{table:uesp_mem_policies}. The \emph{sensitive} code is a code which handles rewriting flash memory, copying data from flash to RAM and setting up memory permissions and is made non-executable after execution. The \emph{MPU Config} refers to MPU configuration registers which are made read-only after memory permissions have been set up and the \emph{SCB config} refers to the \texttt{System Control Block (SCB)} configuration registers.

\begin{table}
\scriptsize
	\centering
	\caption{\uESP Memory Permission Policies}
	\begin{tabular}{p{2.4cm}K{3.9cm}}
		\textbf{Memory} & \textbf{Permissions}\\
		\hline
		Code (sensitive) & RO+XN \\
		Code (other) & RO+X \\
		Data & RW+XN \\
		Peripherals & RW+XN \\
		SCB Config & RO+XN \\
		MPU Config & RO+XN \\
	\end{tabular}
	\label{table:uesp_mem_policies}
\end{table}

With the Cortex-M3's MPU available on \texttt{TI LM3S6965} we can enforce \uESP policies by making use of its support for up to 8 memory regions. Memory regions can cover the full 4 GB address space and come with \emph{size} and \emph{permission} (in the form of XN and data access flags) attributes. Memory regions start addresses must be size-aligned (ie. a 2 KB region must start at an address that is a multiple of 2 KB). We do not make use of the available privilege modes because we do not want to assume OS compatibility with them and as such our permissions apply to both privileged and unprivileged modes. Based on the policies in Table~\ref{table:uesp_mem_policies}, we construct a \uESP configuration for the \texttt{TI LM3S6965} in Tables~\ref{table:ti_lm3s6965_uesp1} and \ref{table:ti_lm3s6965_uesp2} representing settings for \emph{execute-from-flash} and \emph{RAM code relocation} scenarios respectively.

We start with a \emph{default} region, using the lowest region number, which covers the entire address space with \texttt{RW+XN} permissions. If two memory regions overlap on the MPU, region attributes fall back to the region with the highest region number. We can use this feature to limit the number of regions we have to specify and define overlapping regions for exceptions to default \emph{data memory}. \emph{SRAM} and \emph{peripherals} are covered by this default region as well. We define a region for code (covering all of flash memory) with \texttt{RO+XN} permissions and use a higher region as an \texttt{XN} overlay for any \emph{sensitive} code (except the final line which locks the MPU) to be made non-executable after system initialization. 

\begin{table}
\scriptsize
	\centering
	\caption{\texttt{TI LM3S6965} MPU \uESP Settings, \emph{execute-from-flash}}
	\begin{tabular}{p{1.4cm}K{2.1cm}K{1.5cm}K{1.1cm}K{1.1cm}}
		\textbf{Region No.} & \textbf{Description} & \textbf{Perms.} & \textbf{Size}\\
		\hline
		0 & Default & RW + XN & 4 GB \\
		4 & SCB & RO + XN & 64 B \\
		5 & MPU & RO + XN  & 64 B \\
		6 & Code (other) & RO + X & 256 KB \\
		7 & Code (sensitive) & RO + XN & * \\
	\end{tabular}
	\label{table:ti_lm3s6965_uesp1}
\end{table}

For the scenario where code is executed from RAM, we provide identical regions to be placed wherever in RAM the bootloader relocates code to. Note that since \emph{aliased} address ranges need to be covered by the same permissions, this might mean memory regions could need sizes bigger than the actual amount of on-chip SRAM.

\begin{table}
\scriptsize
	\centering
	\caption{\texttt{TI LM3S6965} MPU \uESP Settings, \emph{execute-from-RAM}}
	\begin{tabular}{p{1.4cm}K{2.9cm}K{1.5cm}K{1.1cm}K{1.4cm}}
		\textbf{Region No.} & \textbf{Description} & \textbf{Perms.}& \textbf{Size}\\
		\hline
		0 & Default & RW + XN  & 4 GB \\
		2 & SCB & RO + XN  & 64 B \\
		3 & MPU & RO + XN  & 64 B \\
		4 & Code (other, RAM) & RO + X  & * MB \\
		5 & Code (sensitive, RAM) & RO + XN & * \\
		6 & Code (other, flash) & RO + X & 256 KB \\
		7 & Code (sensitive, flash) & RO + XN & * \\
	\end{tabular}
	\label{table:ti_lm3s6965_uesp2}
\end{table}

In addition to the above we have to consider the following security-sensitive memory regions: \emph{Interrupt Vector Table (IVT)} and \emph{System Control Block (SCB)}. The \texttt{IVT} holds exception vectors such as the stack pointer reset value and start address (loaded upon system reset) as well as interrupt handler addresses. The \texttt{IVT} would be an interesting overwriting target for attackers but by default it lives completely within the lower region of flash memory and as such is covered by the \texttt{RO+X} permissions of our code region. It is possible, however, to relocate the vector table using the \emph{Vector Table Offset Register (VTOR)} in the \emph{System Control Block (SCB)}. If the vector table is relocated to RAM along with other code as part of a bootloader, this is not an issue because it will be covered by the relevant code region. But to protect the system from forcing a malicious relocation as part of an exploit (among other things), we mark the \texttt{SCB} as read-only. If some \texttt{SCB} functionality should be writable during runtime, \uESP uses the MPU's \emph{sub-region} feature which divides a region into 8 equally large sub-regions (provided the region size is at least 256 bytes) that can be disabled individually thus falling back to \emph{default} \texttt{RW+XN} permissions. If so desired, one could merge the \texttt{SCB} and \texttt{MPU} memory regions into a single region using disabled sub-regions to cover any addresses within the range which should have different permissions.

\subsection{$\mu$Scramble Implementation}

The \uScramble diversification transformations were implemented as \emph{LLVM} passes which we describe in the following. 
\paragraph{Register-Preservation Reordering} This transformation is implemented as a \emph{MachineFunctionPass} which obtains the callee-saved registers of a function using \texttt{getCalleeSavedRegs}, shuffles their order using the LLVM PRNG and sets the new order.

\paragraph{Dead Code Insertion} This transformation is implemented as a \emph{MachineFunctionPass} which identifies the final basic block of a given function and generates \emph{dead code}-stubs.
We allow developers to specify what type of \emph{dead code}-stub (\texttt{NOP} or \emph{trap}) they wish to generate with a compiler flag. \texttt{NOP}-stubs consist of a single, repeated, architecture-dependent \texttt{NOP} instruction to ensure minimal gadget usefulness. 
 
		\emph{Trap}-stubs consist of branch instructions to a \emph{violation policy handler}, in our case we use the same handler used for \uSSP violations described below.

\paragraph{Function Reordering} This transformation is implemented as a \emph{ModulePass} which retrieves the current modules function list and shuffles it using the LLVM PRNG. The linker will ensure functions are organized in the randomized order in the produced firmware image. All randomization operations used in \uScramble draw upon the LLVM PRNG which draws upon a developer-supplied \emph{true random} seed. 
Since this PRNG is deterministic this means a given firmware build can be reproduced from the seed as is done by Gionta et al.~\cite{gionta_diversification:2016}.

\subsection{$\mu$SSP Implementation}\label{sec:sspimplement}

We implemented \uSSP as an augmentation of \emph{Zephyr}'s SSP implementation. Since \emph{Zephyr} uses the \emph{GCC} SSP model it already meets \uSSP's compiler-side criteria. On the OS side, it stores a single master canary value as a global variable in \texttt{.bss} and initializes it at boot (as part of the \texttt{\_Cstart} function, after hardware initialization but before the main thread is activated) by drawing from the \texttt{sys\_rand32\_get} API.

We augmented this SSP implementation by adding support for a \emph{terminator-style} canary bitmask, ensuring an OS CSPRNG is available for secure canary generation and implementing a modular canary violation handler. 

\begin{itemize}
	
	\item \textbf{Passive}: The violation handler simply returns to the violating function.
	
	\item \textbf{Fatal}:  The violation handler try to terminate the violating thread and continue running the system.
	
	\item \textbf{Thread Restart}: To properly handle thread restarts we maintain a global list (a \emph{hash table}) of thread restart handlers associated with thread IDs. We require the thread ID be registered together with the restart handler upon thread start and de-registered upon thread termination. Upon invocation of the violation handler, the violating thread's ID is looked up in the restart handler list, the associated restart handler is fetched, the thread in question is terminated and the restart handler is invoked.
	
	\item \textbf{System Restart}:  We invoke the \texttt{sys\_reboot} API with the \texttt{SYS\_REBOOT\_COLD} argument to perform a system restart.

	\item \textbf{System Shutdown}: This approach depends on \emph{SoC} power management subsystem implementations. In the absence of such functionality we default to terminating all running threads.
	
\end{itemize}

\subsection{$\mu$RNG Implementation}

We implemented \uRNG as a driver for the \emph{Zephyr} \texttt{random} API.  \uRNG output can be requested with the \texttt{sys\_rand32\_get} API which \emph{squeezes} the \uRNG \texttt{keccak} object to produce 64 bits (the \emph{rate} minimum) of PRNG output, the upper and lower halves of which are xor-summed together to produce a 32-bit random number as per API specifications.
Since our \uRNG implementation uses SRAM SUVs as its initial entropy source, it is important that this entropy collection takes place as early as possible to reduce SRAM contamination (from code storing variables, using the stack, etc.) as much as possible. We chose to integrate \uRNG initialization in \emph{Zephyr}'s \texttt{\_\_start} routine which is the firmware code entrypoint and used as the \texttt{reset handler} in the ARM Cortex-M's \texttt{vector table}.  Note that almost all RTOSes provide such initialization routine. This ensures SRAM is untouched before our \uRNG initialization code is invoked.

%% file: discussion.tex
\section{$\mu$Armor Evaluation \& Limitations}\label{ch:evaluation} %

\subsection{Overhead Evaluation}

We evaluate the overhead imposed by \uArmor in terms of code size, data size, memory usage, and runtime increases. Code and data size figures represent increases in code (in flash) and constants (in SRAM) respectively. Memory usage increases represents a worst-case SRAM overhead imposition (by use of dynamic data structures) at any point during execution. We instrumented application code to measure runtime performance using a hardware high precision counter. Runtime performance figures represent increases in the number of clockcycles consumed for a given amount of code to run, reported as the average of 25 runs.

We evaluate \uESP, \uSSP, \uRNG separately in Tables \ref{table:uarmor_application_overhead_uesp}, \ref{table:uarmor_application_overhead_ussp} and \ref{table:uarmor_application_overhead_urng} and \uScramble in Tables \ref{table:uarmor_overhead_uscramble_application1}. Note that evaluation results for memory overhead is based on worst-case estimate and the results for runtime overhead are from average of 25 runs. To get an idea of the overhead on realistic applications we chose three sample \emph{Zephyr} IoT applications stressing different subsystems: Additionally, due to space restrictions, we eliminated in Table~\ref{table:uarmor_overhead_uscramble_application1} the results for the data size, memory, and runtime overhead with respect to application and runtime since there was no difference before and after using \uScramble. 

\begin{itemize}

\item \texttt{philosopher}: An implementation of the dining philosophers problem using multiple preemptible and cooperative threads of differing priorities~\cite{zephyr_philosophers:2017}.

\item \texttt{net/echo\_server}: An IPv4/IPv6 UDP/TCP echo server application~\cite{zephyr_echo_server:2017}.

\item \texttt{net/telnet}: IPv4/IPv6 telnet service providing a shell with two shell modules: \texttt{net} and \texttt{kernel}~\cite{zephyr_telnet:2017}.

\end{itemize}

\begin{table}[!tbh]
\scriptsize
	\centering
	\caption{\uScramble Overhead with respect to Applications (A) and Resources (R) in average of 25 variants.}
	\begin{tabular}{p{1.85cm}K{0.75cm}K{0.75cm}|p{1.83cm}K{0.75cm}K{0.75cm}}
		\textbf{App} & \textbf{\% CS\textsuperscript{1} (A)} & \textbf{\% CS \textsuperscript{1} (R)} & \textbf{App} & \textbf{\% CS (A)\textsuperscript{1}  } & \textbf{\% CS\textsuperscript{1}  (R)}   \\
		\hline
		\rowcolors{5}{white}{lightgray}\textbf{Application} & & &\textbf{Test} & & \\
		\hdashline
		\texttt{lift} & 1.5 & 0 &\texttt{cover} & 1.2 & 0  \\
		\texttt{powerwindow} & 2.2 & 0.1 & 	\texttt{duff} & 3 &  \\
		\texttt{} &  &  &  \texttt{test3} & 0.4 & 0.1\\
		
		\hline
		\textbf{Kernel} & & & \textbf{Sequential}& & \\
		\hdashline
		\texttt{binarysearch} & 3.8 & 0 &\texttt{adpcm\_dec} & 1.1 & 0 \\
		\texttt{bitcount} & 2.3 & 0 & \texttt{adpcm\_enc} & 1.3 & 0 \\
		\texttt{bitonic} & 4.2 & 0 & \texttt{ammunition} & 1.1 & 0.1  \\
		\texttt{bsort} & 3.5 & 0 & \texttt{anagram} & 1.8 & 0   \\
		\texttt{complex\_updates} & 1.6 & 0 & \texttt{audiobeam} & 1.6 & 0  \\
		\texttt{countnegative} & 3.4 & 0 &  \texttt{cjpeg\_transupp} & 1.2 & 0  \\
		\texttt{fac} & 4 & 0 & 	\texttt{dijkstra} & 1.9 & 0 \\
		\texttt{fft} & 2.6 & 0 &    	\texttt{epic} & 1.6 & 0 \\
		\texttt{filterbank} & 1 & 0 & 	\texttt{fmref} & 1 & 0  \\
		\texttt{fir2dim} & 1.5 & 0 & \texttt{gsm\_dec} & 1.3 & 0  \\
		\texttt{iir} & 2.2 & 0 & \texttt{h264\_dec} & 0.8 & 0 \\
		\texttt{insertsort} & 1.9 & 0 & \texttt{huff\_dec} & 1.8 & 0 \\
		\texttt{jfdctint} & 1.3 & 0 & \texttt{huff\_enc} & 1.6 & 0  \\
		\texttt{lms} & 1.4 & 0 & 	\texttt{mpeg2} & 1.1 & 0.1   \\
		\texttt{ludcmp} & 0.9 & 0 & \texttt{ndes} & 1 & 0  \\
		\texttt{matrix1} & 3 & 0 & 	\texttt{petrinet} & 0.2 & 0  \\
		\texttt{md5} & 1.9 & 0 & 	\texttt{rijndael\_dec} & 0.3 & 0  \\
		\texttt{minver} & 0.8 & 0 & 	\texttt{rijndael\_enc} & 0.3 & 0 \\
		\texttt{pm} & 0.8 & 0 &\texttt{statemate} & 0.5 & 0 \\
		\texttt{prime} & 1.8 & 0 &  & &   \\
		\texttt{quicksort} & 0.9 & 0 & &  &  \\
		\texttt{recursion} & 4.6 & 0 &  &  &   \\
		\texttt{sha} & 1.8 & 0 &  &  &  \\
		\texttt{st} & 2 & 0 &  &  &   \\
		\texttt{basicmath} & 0.7 & 0 &  &  &   \\

	\end{tabular}

	\small \textsuperscript{1} \emph{Percentage of Code Size (CS) Overhead }\\
	\label{table:uarmor_overhead_uscramble_application1}
\end{table}

We evaluate \uESP, \uSSP, and \uRNG against the above representative applications compiled for the \texttt{TI LM3S6965} with \emph{GCC} as provided by the \emph{Zephyr SDK}. We evaluate \uScramble against a different set of applications since the overhead imposed by \uScramble on a single application is non-deterministic and is dependent to parameters such as the number of functions. Thus we evaluate \uScramble against a set of 50 benchmarks and applications from the \emph{TACLeBench} suite~\cite{taclebench:2016}. \emph{TACLeBench} consists of self-contained programs without external or OS dependencies and is drawn from well-known (embedded) benchmarking suites such as \emph{DSPStone}~\cite{dspstone:2017}, \emph{MRTC WCET}~\cite{mrtc:2017}, \emph{SNU-RT}~\cite{snurt:2017}, \emph{MiBench}~\cite{mibench:2017}, \emph{MediaBench}~\cite{mediabench:2017}, \emph{NetBench} and \emph{HPEC}~\cite{hpec:2017}. The benchmarks in question are drawn from various embedded domains ranging from automotive and networking to security and telecommunications and are sub-divided into \emph{application}, \emph{kernel}, \emph{sequential} and \emph{test} groups implementing realistic applications, small kernel functions, large sequential functions and artificial stress tests respectively.

\begin{table}[!tbh]
	\scriptsize
	\centering
	\caption{\uESP Overhead Evaluation with respect to (wrt) application and resources}
	\begin{tabular}{p{1.8cm}K{1.1cm}K{1.1cm}K{1.1cm}K{1.1cm}}
		\textbf{Wrt. Application} & \textbf{\%Code} & \textbf{\%Data} & \textbf{\%Memory} & \textbf{\%Runtime} \\
		\hdashline
		\texttt{philosopher} & 1.2 & 0 & $\times$ (0 B) & 0 \\
		\texttt{echo\_server} & 0.2 & 0 & $\times$ (0 B) & 0 \\
		\texttt{telnet} & 0.2 & 0 & $\times$ (0 B) & 0 \\
		
		\hline
		\textbf{Wrt. Resources} & & & & \\
		\hdashline
		\texttt{philosopher} & 0 & 0 & 0 & $\times$ \\
		\texttt{echo\_server} & 0 & 0 & 0 & $\times$ \\
		\texttt{telnet} & 0 & 0 & 0 & $\times$ \\
	\end{tabular}
\label{table:uarmor_application_overhead_uesp}
\bigskip
	\scriptsize
\centering
\caption{\uSSP Overhead Evaluation}
	\begin{tabular}{p{1.8cm}K{1.1cm}K{1.1cm}K{1.1cm}K{1.1cm}}
	\textbf{Wrt. Application} & \textbf{\%Code} & \textbf{\% Data} & \textbf{\%Memory} & \textbf{\%Runtime} \\
	\hdashline
	\texttt{philosopher} & 30.5 & 0 & $\times$ (48 B) & 0 \\
	\texttt{echo\_server} & 26.4 & 0 & $\times$ (84 B) & 0.7 \\
	\texttt{telnet} & 27.3 & 0 & $\times$ (84 B) & 0.7 \\
	
	\hline
	\textbf{Wrt. Resources} & & & & \\
	\hdashline
	\texttt{philosopher} & 0.9 & 0 & 0 & $\times$ \\
	\texttt{echo\_server} & 5 & 0 & 0 & $\times$ \\
	\texttt{telnet} & 5 & 0 & 0 & $\times$ 
	\label{table:uarmor_application_overhead_ussp}
\end{tabular}

\bigskip
	\scriptsize
\centering
\caption{\uRNG Overhead Evaluation}
	\begin{tabular}{p{1.8cm}K{1.1cm}K{1.1cm}K{1.1cm}K{1.1cm}}
	\textbf{Wrt. Application} & \textbf{\% Code} & \textbf{\% Data} & \textbf{\%Memory} & \textbf{\%Runtime} \\
	\hdashline
	\texttt{philosopher} & 10.2 & 0.4 & $\times$ (52 B) & 0 \\
	\texttt{echo\_server} & 1.4 & 0.1 & $\times$ (52 B) & 0 \\
	\texttt{telnet} & 1.5 & 0.1 & $\times$ (52 B) & 0 \\
	
	\hline
	\textbf{Wrt. Resources} & & & & \\
	\hdashline
	\texttt{philosopher} & 0.3 & 0 & 0 & $\times$ \\
	\texttt{echo\_server} & 0.3 & 0 & 0 & $\times$ \\
	\texttt{telnet} & 0.3 & 0 & 0 & $\times$ \\
	
\end{tabular}
\label{table:uarmor_application_overhead_urng}

\end{table}

We are not interested in average memory usage overheads but rather in worst case figures because of potentially unacceptable SRAM pressure. Using \emph{stack depth analysis} embedded developers get an indication of the maximum amount of memory used by the stack in their application. We have decided to obtain a \emph{stack depth estimate} in between the usual lower bounds derived from experimental observation and the upper bounds derived from static analysis. This estimate is derived from multiplying the longest identified \emph{call chain} in the program \emph{Control Flow Graph (CFG)} by the overhead imposed by a single canary. Note that we report overhead figures both with respect to the original unprotected application and with respect to total device resources.

Based on the reported figures above, we can conclude code size overheads stay below 5\% with respect to the application for all components except \uSSP and \uRNG and are less than or equal to 5\% with respect to total device resources for all components. Data size, memory usage and runtime overheads all stay well below 1\% both with respect to the application as well as with respect to total device resources. \uSSP code size overheads are clearly the heaviest overhead imposition of all metrics and components. \uSSP introduces roughly 4 instructions of prologue and 5 instructions of epilogue overhead amounting to 36 bytes per protected function. While the average code overhead with respect to the unprotected application is 28.1\%, we can see overhead in terms of total resource pressure remains equal to or below 5\%.

\subsection{Security Evaluation and Limitations}

\subsubsection{$\mu$ESP Security}

\uESP protects against both code injection and code modification by enforcing a separation between code and data memory, forcing an attacker to use a code-reuse payload. By locking the MPU and rendering \uESP and bootloader code non-executable after it has been run, \uESP protects against code-reuse attacks that seek to circumvent \uESP by means of permission-changing payloads or \emph{ret2bootloader} attacks~\cite{goodspeed_sage_advice:2013,zeronightsavr} that seek to rewrite flash memory with attacker-injected code. 

\subsubsection{$\mu$Scramble Security}

To get an idea of the \textit{entropic quality} of \uScramble transformations we performed a \textit{coverage analysis} consisting of taking our selection from the \emph{TACLeBench} suite and generating 1000 different \uScramble-diversified variants for each benchmark. We then harvest all gadgets from each variant using a ROPGadget~\cite{ropgadget:2017} and determined, for each gadget in each variant, in how many other variants the gadget still resides at the same address. We then obtained the average and maximum gadget survival rates. These gadget survival rates give us an indication as to the quality of \uScramble's \emph{coverage} of the target gadget space. The detailed results of this analysis are reported in Table~\ref{table:uscramble_coverage_analysis1}.

\begin{table}[!htbp]
	\scriptsize
	\centering
	\caption{\uScramble Coverage Analysis.}
	\begin{tabular}{p{1.59cm}K{0.8cm}K{0.8cm}|p{1.4cm}K{0.8cm}K{0.8cm}}
		\textbf{Set} & \textbf{Avg. GS\textsuperscript{1}} & \textbf{Max. GS\textsuperscript{1}} & \textbf{Set} & \textbf{Avg. GS\textsuperscript{1}} & \textbf{Max. GS\textsuperscript{1}} \\
		\hline
		\textbf{Application} & & & \textbf{Application} & & \\
		\hdashline
		\texttt{lift} & 2 & 9 &	\texttt{cover} & 29.8 & 143 \\ 
		\texttt{powerwindow} & 1.2 & 10 & \texttt{duff} & 1.6 & 7 \\
		&&&\texttt{test3} & 0 & 0 \\
		\hline
		\textbf{Kernel} & & &\textbf{Sequential} & &  \\
		\hdashline
		\texttt{binarysearch} & 1.4 & 13 & \texttt{adpcm\_dec} & 0.5 & 4 \\
		\texttt{bitcount} & 63.3 & 180 & 	\texttt{adpcm\_enc} & 0.6 & 5 \\
		\texttt{bitonic} & 85.15 & 166 & \texttt{ammunition} & 3.6 & 147 \\
		\texttt{bsort} & 58.5 & 171 & \texttt{anagram} & 0.6 & 4 \\
		\texttt{complex\_updates} & 68.1 & 199 &	\texttt{audiobeam} & 4 & 79 \\
		\texttt{countnegative} & 2 & 11 & \texttt{cjpeg\_transupp} & 32 & 96 \\
		\texttt{fac} & 2.5 & 4 &\texttt{cjpeg\_wrbmp} & 1.4 & 3 \\
		\texttt{fft} & 1.2 & 5 & 	\texttt{dijkstra} & 1.3 & 5 \\
		\texttt{filterbank} & 59.9 & 209 &	\texttt{epic} & 0 & 0 \\
		\texttt{fir2dim} & 70.5 & 210 & 	\texttt{fmref} & 0.5 & 3 \\
		\texttt{iir} & 2.5 & 8 & \texttt{gsm\_dec} & 1.3 & 43 \\
		\texttt{insertsort} & 101.5 & 202 	& \texttt{h264\_dec} & 100 & 181 \\
		\texttt{jfdctint} & 97 & 193 & 	\texttt{huff\_dec} & 18.8 & 91 \\
		\texttt{lms} & 64.2 & 193 & 	\texttt{huff\_enc} & 5.8 & 55 \\
		\texttt{ludcmp} & 55.4 & 166 & 	\texttt{mpeg2} & 0 & 0 \\
		\texttt{matrix1} & 68.8 & 203 &	\texttt{ndes} & 0.6 & 2 \\
		\texttt{md5} & 0.6 & 7 & 	\texttt{petrinet} & 2 & 2 \\
		\texttt{minver} & 1.3 & 5 & 	\texttt{rijndael\_dec} & 2.5 & 16 \\
		\texttt{pm} & 12.5 & 97 &	\texttt{rijndael\_enc} & 3 & 11 \\
		\texttt{prime} & 0.9 & 5 & \texttt{statemate} & 16.4 & 97 \\
		\texttt{quicksort} & 1.5 & 9 &&&\\
		\texttt{recursion} & 91.6 & 188 &&&\\
		\texttt{sha} & 2.3 & 6 &&&\\
		\texttt{st} & 0.8 & 3 &&&\\
		\texttt{basicmath} & 1 & 4 &&&\\

	\end{tabular}
	
	\label{table:uscramble_coverage_analysis1}
	\small \textsuperscript{1} \emph{Gadget Survival}
\end{table}
The results of our analysis demonstrate that the highest average gadget survival rate is $101.5$ (for \texttt{insertsort}) and the highest maximum gadget survival rate is $210$ (for \texttt{fir2dim}) while most remain well below those numbers. This means that in a worst case scenario, a single gadget survives across roughly 10.15\% of variants on average and across 21\% of them at most, requiring brute-force for all other variants. Thus an attacker constructing a code-reuse payload from a given firmware variant cannot expect any gadget in this payload to work beyond 10.15\% of target devices. Attacker also needs to consider that the gadget survival variants of different gadgets do not necessarily intersect. As such, prospects for \emph{gadget chain} survival are even worse and brute-force search space scales with respect to payload length as well.

\subsubsection{$\mu$SSP Security}

The security offered by \uSSP is inherently constrained by the limits of stack canaries: they only protect against stack buffer overflows targeting \emph{stackframe metadata} and not against other types memory corruption vulnerabilities nor against stack buffer overflows targeting local code or data pointers. That being said, \uSSP draws upon an OS CSPRNG (in the form of \uRNG) to generate canaries with 32 or 24 bits of entropy (depending on whether they are configured to be \emph{terminator style} or not) which are, as such, not susceptible to either the insecure randomness issues or the system-side information leaks affecting the original \emph{Zephyr} SSP canaries in the absence of a TRNG. Since the master canary value is refreshed upon system boot, the \emph{system restart}, \emph{Fatal} and \emph{system shutdown} violation policies are not susceptible to bruteforce approaches at all. The \emph{thread restart} policy is susceptible to bruteforce attacks. All approaches, however, raise at least an alert upon invocation. We believe that given our attacker model, 32 bits of entropy is sufficient for a bruteforce attack to be infeasible.

\subsection{Limitations}

In the following, we discuss limitations of our approach and the prototype implementation.
First, \uArmor requires either a \emph{(modified) Harvard} or \emph{Von Neumann} CPU featuring an MPU with hardware ESP support. We believe that this limitation is within reasonable bounds for our system to be considered \emph{hardware agnostic}, since it only excludes older Von Neumann architectures for which there is currently no way to enforce low-overhead ESP.

Second, the \uScramble component only diversifies \emph{code memory} and only provides \emph{per-device diversity}: it does not diversify \emph{data memory} nor does it diversify on a \emph{per-boot} or \emph{per-application run} basis.
Finally, the \uRNG component requires on-chip SRAM or otherwise requires an alternative source of initial entropy. 

\section{Related Work}
A relevant stream of work has explored software diversification. For example, AVRAND~\cite{avrand} is a boot-time software only diversification scheme that randomizes binary code at a per-page granularity level. The AVRAND imposes an average code size increase of 20\% and requires sizable meta-data storage in EEPROM. Note that AVRAND requires re-flashing flash memory upon randomization which reduces embedded device lifespan. 
MAVR is a boot-time diversification scheme~\cite{mavr} for UAV control system. However, MAVR is not hardware agnostic since it requires (costly) hardware modifications. Note that  both MAVR and AVRAND require re-flashing flash memory upon randomization which reduces embedded device lifespan.
Abbasi et al.~\cite{uShield} introduced $\mu$Shield , a CFI system for embedded COTS binaries with configurable protection policies to cope with limited resources in embedded systems. However, $\mu$Shield addresses high-end embedded devices. 
Another stream of work explored firmware integrity verification for embedded systems. For example Doppelganger~\cite{cui2011defending} protects embedded systems from firmware modification via software symbiotes, a specific integrity verification trap which is invoked every time it gets executed. However, Doppelganger can only verify the static part of the firmware. 

Finally, Clements et al. recently proposed EPOXY~\cite{clements2017protecting}, to protect bare-metal deeply embedded systems. There are several key limitations to the EPOXY~\cite{clements2017protecting} work that set it apart from \uArmor.
First, \uArmor targets a class of deeply embedded systems not addressed by EPOXY: those running an OS. EPOXY was designed to protect single-stack bare-metal embedded systems and is not suitable for deployment on multi-stack RTOS systems without significant re-engineering (e.g., such as with regards to scalability of safe-stack overhead, integrating privilege overlays transparently with OS mechanisms).

Second, EPOXY privilege overlay security model could be broken by an attacker returning to any legitimate call-site where the privilege elevation handler is invoked. Consider, for example, the Algorithm~\ref{algo}:\\

\begin{algorithm}
	\caption{EPOXY Privilege Overlay Issue}\label{algo}
	\begin{algorithmic}[1]
		
		\STATE Begin REQUEST PRIVILEGED EXECUTION
		\STATE Save Register and Flags State
		\IF {In Unprivileged Mode} 
		\STATE Execute SVC 0xFE (Elevate Privileges)
		\ENDIF
		\STATE Restore Register and Flags
		\STATE Execute Restricted Operation
		\STATE Set Bit 0 of Control Register (Reduces Privileges)
		\STATE END
		
	\end{algorithmic}
	
\end{algorithm}

Here an attacker could return to line 2 where supervisor call (SVC 0xFE) would invoke the custom EPOXY SVC\_Handler (see related EPOXY code at~\cite{epoxysrc}). The handler would check whether the interrupt source was correct regardless of unprivileged control-flow leading up to it and as such would elevate privileges allowing the attacker to execute the restricted operation in line 7. Code diversification would not help much because many restricted operations are located at static or easily obtainable addresses (e.g., extracted from the IVT).

Third, EPOXY code diversification approach is less fine-grained than that of \uArmor since it only randomizes function order but does not diversify function sizes (and thus leaves gadget offsets with respect to function addresses intact) nor does it reorder register preservation code. As such, EPOXY code diversification is more vulnerable to information leaks than \uArmor.

Finally, unlike \uArmor, EPOXY does not protect against ret2bootloader style attacks~\cite{francillonharvardinjection,zeronightsavr,goodspeed_sage_advice:2013} because it leaves sensitive bootloader code that cannot be relocated executable after system initialization As a result, an attacker can return to potentially sensitive code areas allowing for re-flashing code memory or disabling the MPU.

%% file: conclusion.tex
\section{Conclusions and Future work}

In this paper, we have presented the first quantitative study of exploit mitigation adoption in a representative selection of embedded operating systems, showing the embedded world (deeply embedded system in particular) to significantly lag behind the general-purpose world. The resulting ease of exploitation of memory corruption vulnerabilities and notoriously prolonged vulnerability exposure windows in the embedded world are cause for concern. We have presented how hardware and OS limitations and performance constraints contribute to an imposing series of constraints for developers of embedded exploit mitigations to overcome. To address this situation we have presented $\mu$Armor, an exploit mitigation baseline design for deeply embedded systems. We have shown that $\mu$Armor holds up favorably in terms of overhead imposition and offered security.

We see two main trajectories for future work on embedded binary security: \emph{long-term} and \emph{short-term} solutions. The former trajectory aims to develop robust techniques tackling the problem at the root and requires changes along the whole development chain. Examples are embedded-oriented safe languages and secure and scalable patching solutions.

The short-term trajectory should aim to develop solutions which reduce the impact of embedded memory corruption vulnerabilities and which can be rapidly adopted.

 Finally, we should also seek for more advanced mitigations for embedded systems in order to continue raising the bar and close the mitigation security gap between general-purpose and embedded systems.

\section*{Acknowledgment}

The work of first and third author was supported by
the European Research Council (ERC) under the European Union's
Horizon 2020 research and innovation programme (ERC Starting
Grant No. 640110 BASTION). In addition, this work was 
supported by the German Federal Ministry of Education and Research (BMBF Grant 16KIS0592K HWSec)
and the Intel Collaborative Research Institute ``Collaborative Autonomous \& Resilient Systems'' (ICRI-CARS). 
The work of second and fourth authors has been supported by the NWO through the SpySpot project (no.628.001.004).